\documentclass[reprint,aps,amsmath,amssymb,prb,superscriptaddress,longbibliography%
]{revtex4-2}
\usepackage{dcolumn} 
\usepackage{graphicx} 
\usepackage{graphicx,subcaption}
\usepackage[version=3]{mhchem}
\usepackage{color,soul}
\usepackage{multirow}
\usepackage{xcolor}
\usepackage{bm}
\usepackage{comment}
\usepackage{float}
\usepackage{amsmath}
\usepackage{physics}
\usepackage{accents} 
\usepackage{xcolor}
\usepackage{bm}
\usepackage{comment}
\usepackage{blkarray}
\usepackage{hyperref}
\usepackage{xr}

\newcommand{\jln}[1]{{\color{black}{#1}}}

\newcommand{\dt}[1]{\accentset{\approx}{#1}}

\newcommand{\hH}{{{\hat{H}}}}

\newcommand{\hbr}{{{\hat{\bm r}}}}
\newcommand{\hbd}{{{\hat{\bm d}}}}

\newcommand{\hbp}{{{\hat{\bm p}}}}
\newcommand{\hbGm}{{{\hat{\bm \Gamma}}}}

\newcommand{\bR}{{{\bm R}}}

\newcommand{\bGm}{{{\bm \Gamma}}}

\newcommand{\bPi}{{{\bm \Pi}}}

\newcommand{\bQ}{{{\bm Q}}}

\newcommand{\br}{{{\bm r}}}

\newcommand{\bx}{{{\bm x}}}
\newcommand{\bP}{{{\bm P}}}
\newcommand{\bp}{{{\bm p}}}
\newcommand{\ba}{{{\bm a}}}

\newcommand{\bk}{{{\bm k}}}

\newcommand{\bq}{{{\bm q}}}
\newcommand{\bl}{{l}} 
\newcommand{\bd}{{{\bm d}}}

\newcommand{\bhr}{{\hat{{\bm r}}}}
\newcommand{\bhp}{{\hat{{\bm p}}}}
\usepackage[margin=1in]{geometry}

\begin{document}
\title{A Phase Space Electronic Structure View of The Solid State} 

\author{Nadine C. Bradbury}
\email{nadinebradbury@princeton.edu}
\author{Linqing Peng}
\author{Joseph E. Subotnik}
\email{subotnik@princeton.edu}
\affiliation{Department of Chemistry, Princeton University, Princeton, NJ USA}
\date{\today}

\begin{abstract}
    We develop a phase space electronic structure view of the  solid state, where electronic bands  are parameterized by both nuclear position  $\bQ$ and nuclear momentum $\bPi$ (rather than the standard practice of only $\bQ$).   Within this phase space, non-Born-Oppenheimer electronic structure context,  we prove a nuclear wavevector ($\bq$)-dependent version of Nafie's equality relating the change in electronic momentum as a function of nuclear momentum, $\partial \left<\bp\right>/\partial \bPi_q$, to the  change in electronic position as a function of nuclear position, $\partial \left<\br\right>/\partial \bQ_q$. Furthermore, we show how information  about electron-phonon interactions, specifically nuclear-induced electronic momentum, can be extracted from simple band structure calculations --  without Berry curvature calculations. 
\end{abstract}
\maketitle


Understanding electron-phonon couplings and momentum transfer is critical for modeling a range  of important phenomena, including heat-capacity, heat-transport, electronic conductivity and superconductivity.  Very often, so-called ``adiabatic'' contributions to the electron-phonon couplings are thought to dominate in solids, especially where the  important nuclear motion has a low frequency  and the electronic motion occurs on a much faster time scale.  That being said, ``non-adiabatic'' contributions also arise, e.g. in the thermal-hall effect\cite{Strohm2005} in metals or phonon-hall effect in insulators,\cite{Saito2019,Zhang2021} or when vibrational modes are strongly coupled to magnons (with nearly degenerate spin states). Indeed, there are many examples where electronic or spin dynamics   proceed on similar timescales\cite{Miglio2020} within phonon dynamics, and there is a non-trivial exchange of momentum  between the two subsystems.

Within the molecular realm (as opposed to the solid state realm), treating  such momentum transfer is quite tractable. For instance, the strong coupling of nuclear angular momentum to both electronic orbital and spin momentum leads to measurable spin-rotation splittings in small molecules  and $\Lambda$-doublings in diatomics\cite{hill1928quantum,VanVleck1929}.  
Using the fact that the  Hilbert space is relatively small  
theory has long been able to predict these energy gaps quite well (and to a very high resolution).\cite{tarczay2010first,lefebvre2004spectra} 

That being said, however,  the situation is much more difficult within the solid state realm, where momentum exchange also occurs but where one cannot apply a brute-force solution. For example, 
the presence of a nonzero electronic spin in a given material can lead to different energies for a pair of chiral phonons at the Brillouin-zone (BZ) center. When the spin is experimentally driven, this effect is the Einstein de Haas effect,\cite{einstein_dehaas_1915b,Ganzhorn2016_einsteindehaas,Wallis2006_EdH_in_NiFe}; the inverse case (where the phonon is driven) is  the Barnett effect.\cite{Basini2024} Modeling such effects is quite difficult and remains a major goal of modern condensed matter physics\cite{SchmidtLemeshko2015}.

In order to model the above effects, one major culprit is the fact that  modern electronic structure calculations inevitably begin by diagonalizing the electronic hamiltonian with frozen nuclei, a step which formally results in 
many electronic and nuclear states that must still be coupled together according to the standard Born-Huang framework.  
{\em Ab initio} theory for solids has been able to make progress by focusing on the ground state and $(i)$ including nuclear berry curvature terms  proportional to the nuclear velocity within the interatomic force constants before  $(ii)$ evaluating harmonic phonon energies, which are now split by spin.\cite{Bonini2023, Ren2024}  Others have approached the same non-adiabatic problem through an exact factorization framework,\cite{Cohen2025_exact_factorization,Abedi2010_exact_fctorization} which is able to connect the perspective of the Born-Huang (BH) framework to the self consistent diagrammatic Hedin-Baym equations of Giustino.\cite{Giustino2017} 
Others have  used a nuclear `diabatic'-BH perspective.\cite{Mohanty2019}
The list of approaches above is certainly incomplete and the question of how best to model electron-phonon coupling in solids remains incomplete--there is indeed a great deal of information missing when one starts from a Born-Huang frozen nuclei calculation.

To that end, in this letter, while eventually aiming to avoid the Born-Oppenheimer approximation entirely,  we will study solids  through a novel, quite orthogonal phase space electronic framework inspired by the work of Shenvi\cite{Shenvi2009-jcp}. Within such a framework,   nuclei are transformed  into classical symbols using a Wigner transformation and the electronic Hamiltonian $\hH_{PS}(\bR,\bP)$ depends on both nuclear position $\bR$ and momentum $\bP$.
Previous studies in molecules  have demonstrated that such an approach can  yield accurate and meaningful spectroscopic properties, ranging from vibrational circular dichroism,\cite{Duston2024_vcd} non-resonant Raman optical activity,\cite{Tao2026_roa}, lambda-doubling splittings\cite{Peng2026_lambda}, and the nuclear angular momentum induced splittings found in the rotational spectra of radical systems.\cite{Peng2026_spinrot}
Although Refs. \citenum{Duston2024_vcd,Tao2026_roa,Peng2026_lambda,Peng2026_spinrot} have all invoked the translational and rotational symmetry of free space when constructing $\hH_{PS}$, we will show here that the same approach can also be used to study solid-state materials when properly interpreted and with minimal cost (e.g. without the need to calculate berry curvatures). Thus,  we hypothesize that  the diagonalization of $\hat{H}^{PS}_k(\bR,\bP)$ will allow us to study  new and rich physics for realistically large {\em ab initio} systems in the  near future.

\bigskip


\paragraph*{Standard electron-phonon couplings through a Born-Huang framework:}
Before discussing phase space electronic structure theory,  let us briefly review the standard Born-Huang approach to electron-phonon couplings. We begin  with a Hamiltonian that treats electrons with second quantization and distinguishable nuclei with first quantization. The position of unit cell $\ell$ is denoted $\bR_{\ell}$, where $\bR_\ell = \ell_1\ba_1 + \ell_2\ba_2+\ell_3\ba_3$ for primitive lattice vectors $\ba$.\cite{Baroni2001_DFPT}
The position of atom $A$ in unit cell $\ell$ is denoted $\bR_{A\ell}$. Bold font denotes vectors, and hats are used for electronic operators.
Though not common with solid-state calculations, we begin with a Wigner transformation over the nuclear degrees of freedom where we assume we have $N$ unit cells:
\begin{align}
\hH(\bR,\bP) &= \sum_{A,\ell=1}^N \frac{\bP^2_{A\ell}}{2M_{A\ell}} + \frac{\bhp^2}{2m_e} + V(\bhr,\{\bR_{A \ell}\})  \label{eq:H_periodic_full_wigner}
\end{align}
Within this framework, the  nuclear kinetic energy becomes a symbol (i.e.  parameter); for nuclei arranged in a periodic lattice, Eq.  \ref{eq:H_periodic_full_wigner} is routinely solved using Bloch theory, producing energies  $\epsilon_{n\bk}(\bR,\bP)$ and wavefunctions  $\psi_{n\bk}(\br;\bR) = e^{i\bk \cdot \br} u_{n\bk}(\br;\bR)$ that are the eigenvalues of $H_k$:
\begin{align}
&    \hH_{\bk}(\bR, \bP) = \sum_{A\ell} \frac{\bP_{A\ell}^2}{2M_{Al\ell}} +  \frac{(\bhp + \hbar\bk)^2}{2m_e} + V(\bhr,\{\bR_{A \ell}\})\label{eq:Hk} 
\end{align}
 When the atoms are displaced, and strict periodicity is lost, one can still extrapolate
  $\epsilon_{n\bk}(\bR,\bP)$ and $\ket{\psi_{n\bk}(\bR)}$ by continuity, even though the meaning of the index  $k$
becomes limited. Note that $u_{n\bk}(\br+\ba;\bR) = u_{n\bk}(\br;\bR)$ only for a periodic lattice. 

Now, let us  define phonon coordinates $(\bQ,\bPi)$ a-la monochromatic phonon density functional perturbation theory (DFPT)\cite{Baroni2001_DFPT}:
\begin{align}
    \bQ_{A\bq} &= \frac{1}{\sqrt{N}} \sum_{\bR_l} e^{- i \bR_l\cdot \bq} (\bR_{A\bl} - \bR_{A\bl}^0)
\end{align}
 In this work, we have assumed the sum over $\bR_\ell$ is finite (and thus only some values of $\bk$ and $\bq$ are allowed), and $N$ refers to the number of unit cells. Consequently, items like $\langle \bhr\rangle$ are extensive in $N$.  Next, let us further define $\bPi$ to  be the momentum conjugate to $\bQ$. Noting that Wigner-Weyl transforms commute with linear canonical changes of coordinates (as reviewed in SI section 1A
),  the full electron-phonon Hamiltonian can then be recovered by a Weyl transform over the Hamiltonian in the adiabatic basis, $\hH = W^{-1}\left(\hH_W\right)$, where
\begin{align}
    \hH_W =  \sum_{n\bk} \ket{{\psi}_{n\bk}(\bQ)} & \Biggl( 
    \sum_A \bPi_{A -\bq} \cdot \bPi_{A \bq} /2M_A   
    \nonumber \\
    & + \epsilon_{n\bk}(\bQ)
    \Biggr) \bra{\psi_{n\bk}(\bQ)} \label{eq:H_W_fullbasis}
\end{align}

An expansion of the electronic bands into a basis of plane waves, $\left\{ e_y \right\}$ leads to  coefficient matrices $Z_{y,n\bk}(\bQ) = \bra{e_y}\ket{\psi_{n\bk}(\bQ)}$. We can rigorously (unitarily) transform the Hamiltonian in  Eq. \ref{eq:H_W_fullbasis}  to the $\left\{e_y\right\}$ basis by working with the Moyal star product
\begin{align}
    \hat{H}'_W &= Z^{\dagger}(\bQ) * \hH_W(\bQ,\bPi) * Z(\bQ)  
\end{align} 
which yields (ignoring the
second order term $\hat{\zeta} \equiv \sum_{Aq} \frac{1}{2M_A} \frac{\partial Z^{\dagger}}{\partial \bQ_{A-\bq}} \frac{\partial Z}{\partial \bQ_{A\bq}}$)
\begin{align}    
    \hat{H}'_W & \approx \sum_{A\bq} \frac{\bPi_{A-\bq} \cdot \bPi_{A\bq}}{2M_A}   - i \hbar \frac{\bPi^{A\bq} \cdot \bd^{A-\bq}}{M_A} +\epsilon(\bQ) \label{eq:H_QShenvi}
\end{align}
Eq. \ref{eq:H_QShenvi} is the standard Born-Huang expansion for a periodic solid after a Weyl transform over $\bPi$. 
To derive Eq. \ref{eq:H_QShenvi}, note that  
the Moyal star product truncates at  second order in $\hbar$ and we have introduced the nuclear berry connection $d^A_q \equiv Z^\dagger \frac{\partial Z}{\partial \bQ_{A\bq}}$, or derivative couplings as they are known in chemistry. 

So far we have used the language of Wigner-Weyl transforms, but the result is equivalent to standard DFPT theory, provided that we  expand 
$\epsilon(\bQ)$ to second order (i.e. a harmonic approximation).
The matrix elements of $\hat{\bd}^A$ are the electron-phonon matrix elements\cite{Giustino2017} divided by the band energy gap.
\begin{align}
    d^{Aq}_{n\bk',m\bk}  = N \frac{\bra{u_{n\bk'}} \delta \hat{W}^{A\bq}  \ket{u_{m\bk}}}{\epsilon_{n\bk+\bq}-\epsilon_{m\bk}}\delta_{\bk',\bk+\bq} \label{eq:dAq}
\end{align}
Here, $\ket{u_{n\bk}}$ is the cell-periodic part of the  Bloch state $\ket{\psi_{n\bk}}$, and $\delta \hat{W}^{A \bq}$ is defined as (just as for DFPT\cite{Baroni2001_DFPT})
$\delta \hat{W}^{A\bq}(\br) \equiv \sum_{\ell} -\frac{\partial \hat{V}_{Al}(\br)}{\partial \br} e^{-i(\br-\bR_l)\cdot\bq}.$
From the delta function in Eq. \ref{eq:dAq}, it is clear that the elements of $\hat{\bd}^A$ conserve crystal momentum.

\bigskip

\paragraph*{Momentum Conservation and Nafie's Equality:}

Within the realm of solid-state electronic structure, the focus is usually on crystal momentum, rather than canonical momentum. That being said,  for a system in free space {\em or} for a system with periodic boundary conditions,  Noether's theorem holds and there is linear momentum conservation\cite{frenkel_and_smit}. Moreover, within the molecular world (where there is always a focus on momentum conservation),  many decades ago, Nafie proved that, if one takes 
\begin{align}
\hH_{\rm Nafie}(\bR,\bP) = \hH_{el} - i\hbar \sum_A \frac{\bP^A \cdot \hbd^A}{M_A}  
\label{eq:define_Hnafie}
\end{align}
as a perturbative hamiltonian to account for electron-phonon couplings, and one further diagonalizes  $\hH_{\rm Nafie} \ket{\psi} = E \ket{\psi}$, there is a remarkably simple result,  
$m_e \, \frac{d}{dt} \left<\psi \middle|  \hbr \middle|\psi \right>  = \left< \psi \middle| \hbp \middle| \psi \right>$   which can be rewritten as:

\begin{align} 
\frac{\partial\langle \psi | \hat \br | \psi \rangle}{\partial \hat \bR} &= \frac{M_A}{m_e}\frac{\partial\bra{\psi} \hat \bp \ket{\psi}}{\partial \bP} \Biggl|_{\bP = 0}\label{eq:Nafie_equality_original}
\end{align}

As we will show, this molecular result has an analogue in the solid-state world.  For a given displacement \jln{wave-vector}, $\bq$, let us define the (non-observable, complex) $\bq-$dependent \jln{operator 
\begin{align}
    \hbp_{\bq} = \frac{1}{2}\left(e^{-i\bq \cdot \hbr} \hbp + \hbp e^{-i\bq \cdot \hbr} \right)
\end{align}
similar to how others have done for the electronic current.\cite{Dreyer2018}. We then define an additional operator,
\begin{align}
    \zeta_{\bq} =  \left( \frac{1 - e^{-i \bq \cdot \hbr}}{i|\bq|} \right),
\end{align}
such that
\begin{align}
    \left[\hH_W, \zeta_\bq\right] &= -i\hbar \frac{\bhp_\bq\cdot \bm{n}_\bq}{m_e}. 
\end{align}
for the unit vector $n_\bq = \bq/|\bq|$.
}

In the SM, we show that the following analogue to Nafie's result holds:
\jln{
\begin{align}
            \mbox{Re} \left\{\frac{\partial\langle \psi_{n\bk} | \hat \zeta_{\bq}  | \psi_{n\bk} \rangle}{\partial \bQ_{A\bq}}\right\}= \frac{M_A}{m_e}\mbox{Re} \left\{\frac{\partial \bra{\psi_{n\bk}} \hat \bp_{\bq} \cdot  \bm{n}_\bq\ket{\psi_{n\bk}}}{\partial \bPi_{A,-\bq}} \right\}\label{eq:Nafie_equality}
\end{align}
}

At the end of the day, the left hand side of  Eq.  \ref{eq:Nafie_equality} can be evaluated using the eigenstates of $\hH_{el}$ (which do not depend on $\bP$) and the result easily reduces to
\begin{widetext}
\begin{align}
    LHS&=      
       - N  \sum_{\bk',c\neq n}  
\Im{\frac{\bra{u_{n\bk}} 
\delta \hat{W}^{A\bq}
    \ket{u_{c \bk'}}\bra{u_{c \bk'}}
    \jln{\bm{n}_\bq \cdot(\bhp + \hbar\bk) - \hbar\bq/2}
    \ket{u_{n\bk}}}{(\epsilon_{c \bk'}-\epsilon_{n \bk})^2}} 
\delta_{\bk',\bk-\bq}
\nonumber
   \\
   & +
 N  \sum_{\bk',c\neq n}  
\Im{\frac{\bra{u_{n\bk}} 
    \jln{\bm{n}_\bq \cdot(\bhp + \hbar\bk) - \hbar\bq/2}
    \ket{u_{c \bk'}}\bra{u_{c \bk'}}
    \delta \hat{W}^{A\bq}
    \ket{u_{n\bk}}}{(\epsilon_{c \bk'}-\epsilon_{n \bk})^2}} 
\delta_{\bk',\bk+\bq} \label{eq:Nafie_LHS} 
\end{align}
\end{widetext}

The expression in Eq.  \ref{eq:Nafie_LHS} should be familiar to most chemical and condensed matter physicists, in so far as the  form is clearly a berry  curvature expression for the atomic polar tensor (at $\bq=0$).\cite{Resta2023} To demonstrate that the right hand side of Eq. \ref{eq:Nafie_equality} also  reduces to Eq. \ref{eq:Nafie_LHS}, we require a momentum dependent hamiltonian (not just $\hH_{el}$) and 
we need only  substitute  $\hat{H}'_W$  in Eq. \ref{eq:H_QShenvi} for the analogue of $\hH_{\rm Nafie}$ in Eq. \ref{eq:define_Hnafie}$ $ and invoke Eq. \ref{eq:dAq}.

Interestingly, as shown in the SM, the \jln{RHS} equality in Eq. \ref{eq:Nafie_equality} can be viewed as a symmetry of a $\bq-$dependent mixed nuclear-electronic curvature-like tensor. Namely, let us define
\begin{align}
    \Omega^{\bq}_{n\bk}(\alpha_{\bq},\beta) &\equiv 
     \frac{1}{2}
    \mbox{Im} \Biggl[ \left<
    \frac{\partial}{\partial \alpha_{-\bq}} \psi_{n \bk} \middle|  
    e^{-i\bq\cdot\br}  \frac{\partial}{\partial \beta} \psi_{n \bk} 
    \right> \\
    & -
    \left<
    \frac{\partial}{\partial \beta} \psi_{n \bk}
     \middle| 
     e^{-i\bq\cdot \br}  
     \frac{\partial}{\partial \alpha_{\bq}} \psi_{n \bk}
    \right>
    \Biggr]
  \label{eq:Berry}
\end{align}
Our proposed extension of Nafie's equality is then equivalent to the symmetry
\begin{align}
    \Omega^\bq_{n\bk}(\bQ_{A\bq},\tilde{\bk}) = 
    \frac{-M_A \hbar}{m_e} \Omega^\bq_{n\bk}(\bPi_{A-\bq},\br) \label{eq:Curvature_equality}.
\end{align}
awhich resembles a Maxwell relation and establishes an equality relating the position and momentum operators very much like the length/velocity gauge equalities that are well known in spectroscopies; Eq. \ref{eq:Curvature_equality} is valid for any two electronic states and anywhere in the BZ, \jln{where analogous to the standard quntum geometry result that $\bhr\to i\partial_\bk$ for a single $\bk$ point of, we now use a heuristic definition for $\zeta_{\bq} \to e^{-i\bq\cdot\bhr}\partial_{\tilde{\bk}}$over a range of $\bk$ points, and reduces to the same $i\partial_\bk$ expression in the limit $\bq \to 0$}.

\paragraph*{An approximate PS theory:} Drawing on the approach  above inspired by Nafie, the ansatz of modern phase space electronic structure theory is that we can go beyond standard BO theory and learn directly about electron-phonon interactions  by diagonalizing an eletronic operator parametrized by both $\bR$ and $\bP$ (without calculating a berry curvature).
The key insight is that we can
use symmetry to approximate the derivative coupling $\hat{\bf d}^{A\bq}$ in Eq. \ref{eq:H_QShenvi} in a fashion that does not require the energy denominator in Eq. \ref{eq:dAq}. Given the fact that $\hat{\bf d}^{A\bq}$ is often akin to a local electronic momentum, 
as evidenced by the equality 
\begin{align}
        i\hbar \sum_A d^{A, \bq=0}_{n\bk,m\bk'} = \bra{\psi_{n\bk}}\bhp\ket{\psi_{m\bk'}}
        \label{eq:sumd}
\end{align}
we hypothesize that a good approximation to $\hat{\bf d}^{A\bq}$  is:
\begin{align}
\hat{\Gamma}^{A\bq}(\br)   =  \frac{1}{2i\hbar} \biggl(&\sum_{\ell} \theta_{A,\ell}(\br)  e^{i\bR_\ell\cdot\bq} \bhp + \bhp e^{i\bR_\ell\cdot\bq} \theta_{A,
\ell}(\br) 
\biggr)
\label{eq:Gamma_new}
\end{align}
Here, we have defined a partition of unity function $\theta_{Al}(\br)$ which is nonzero only when $\br$ is near $\bR_{Al} = \bR_\ell + \tau_A$:
\begin{align}
    \theta_{A\ell}(\br) = \frac{e^{-(\br-\bR_\ell -\tau_A)^2/\sigma^2}}{\sum_A\sum_{\ell'} e^{-(\bx-\bR_\ell'-\tau_A)^2/\sigma^2}}
\end{align}
$\sigma$ is a broadening length, e.g.  $\sigma \approx 1 $ bohr. Note that the analogue of Eq. \ref{eq:sumd} is  the fact that $ i\hbar \sum_A \hat{\Gamma}^A_0 =\bhp $. Note also that 
Eq. \ref{eq:Gamma_new} naturally satisfies  crystal-momentum conservation condition insofar as $\left( \hat{\bf \Gamma}^{A \bq}\right)_{\bk'\bk} \propto \delta_{\bk',\bk+\bq}$
(as in Eq. \ref{eq:dAq} for $\hat{\bf d}^A$).

Finally, if we seek to diagonalize a 
 potential that is periodic over the unit cell with some nuclear momentum information, we submit that the relevant phase space electronic structure Hamiltonian in the periodic gauge is of the form:
\begin{align}
    H^{PS}_{\bk}&(\bR,\bP) = \sum_A \frac{(\bP_{A} - i\hbar \hat{\bf \Gamma}^{A,\bq = 0}_{\bk})^2}{2M_A} \nonumber \\
    &+  \frac{(\bhp + \hbar\bk)^2}{2m_e} + V(\br,\bR) \label{eq:HPS_q0}
\end{align}
Let us denote the eigenvalues of $\hH^{PS}_k$ as $\epsilon^{PS}{(\bR,\bP)}$ and the wavefunctions as $\psi^{PS}$. Importantly,  these eigenstates exhibit electronic momentum \emph{beyond} the crystal momentum, 
\begin{align}
    \bra{u^{PS}(\bR,\bP)} -i\hbar \partial_\br \ket{u^{PS}(\bR,\bP)} \neq 0
\end{align}
when $\bP \ne 0$.

\begin{figure}[h]
    \centering
    \includegraphics[width=\linewidth]{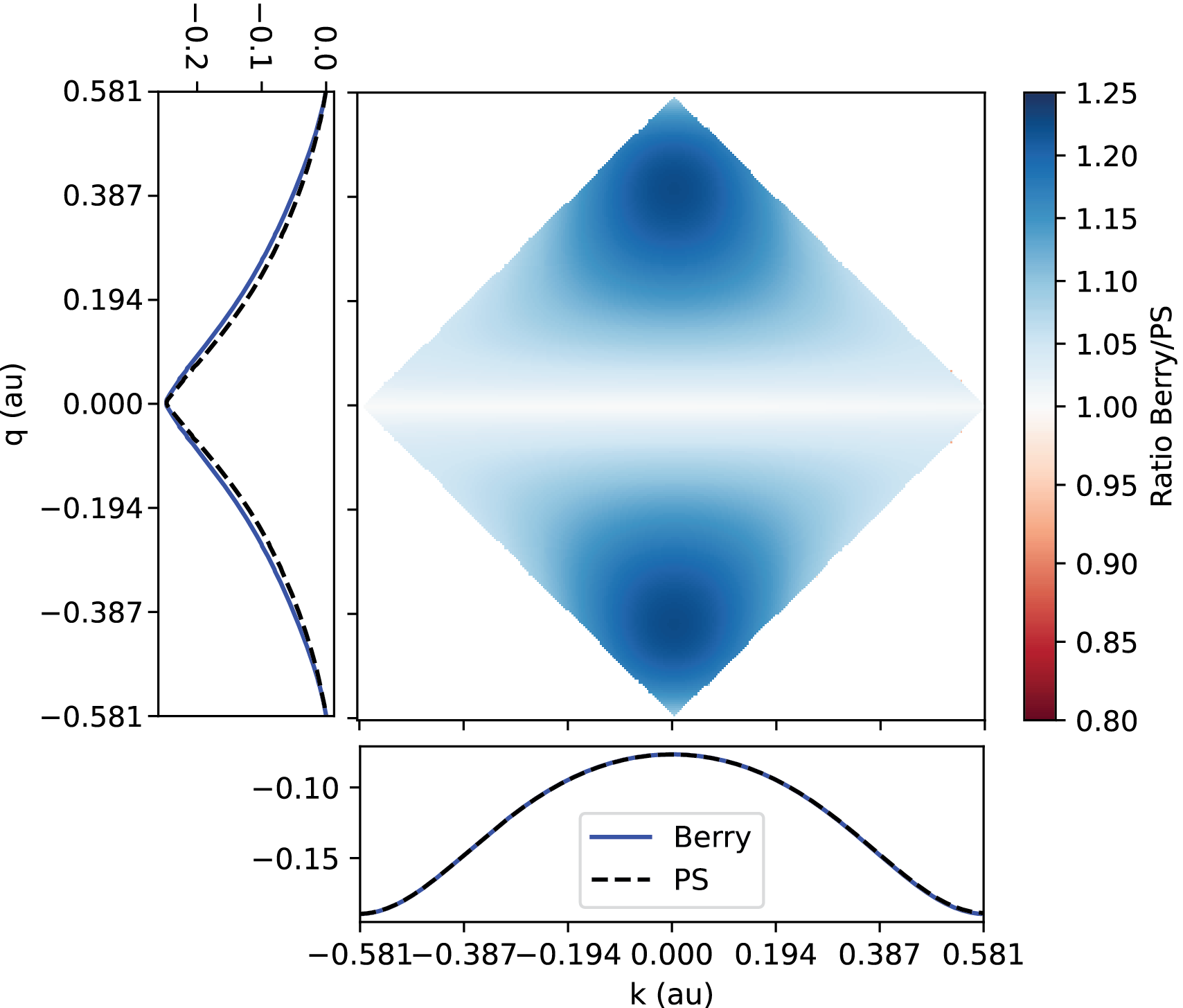}
    \caption{In the centrally located colored density heat map,  we plot the ratio of the LHS  in Eq. \ref{eq:Nafie_LHS} and RHS in Eq. \ref{eq:nafie_RHS_PS} for every combination of $k$ and $q$ that remains in the first BZ (average value 1.16). This ratio is exactly unity for an exact version of Nafie's theorem using the derivative coupling as the electron-nuclear coupling element (as in Eq. \ref{eq:define_Hnafie}).  To the left of the heat map, we  plot the BZ-average (i.e. average over $k$) of Nafie's equality as a  function of $q$. Below the heat map, we fix $q=0$  and plot  the ratio  as a function of $k.$ The fact that the LHS/RHS ratio is so close to unity validates our approximation for $\hbGm$ within a  phase space electronic structure approach.\vspace{-5mm}}
    \label{fig:Nafie_ratio}
\end{figure}

\bigskip
\paragraph*{Validation of Nafie's equality:}
One means to verify that the formalism above has  physical meaning is to demonstrate that Nafie's equality (in Eq. \ref{eq:Nafie_equality}) is approximately retained.
To that end, 
in Figure \ref{fig:Nafie_ratio}, we compare the left and right hand sides of Eq.\ref{eq:Nafie_equality} within the context of a phase space electronic structure theory for a simple one-electron Hamiltonian.  Thus, for the LHS, we  evaluate  Eq. \ref{eq:Nafie_LHS} directly; for the RHS, 
we need only  substitute  $\hat{\Gamma}^{A\bq}$ for $\hat{\bd}^{A \bq}$ in $\hat{H}'_W$  in Eq. \ref{eq:H_QShenvi}
and thereafter differentiate. In short, we will compare  Eq. \ref{eq:Nafie_LHS} above with Eq. \ref{eq:nafie_RHS_PS} below:
\begin{widetext}
\begin{align}
    RHS &=  - N \frac{\hbar}{m_e} \sum_{\bk',c\neq n} \Im{\frac{\bra{u_{n\bk}}\dt{\Gamma}^{A,\bq} \ket{u_{c \bk'}}\bra{u_{c\bk'}}\jln{\bm{n}_\bq \cdot(\bhp + \hbar\bk) - \hbar\bq/2} \ket{u_{n\bk}}}{\epsilon_{c\bk'}-\epsilon_{n\bk}}} \delta_{\bk',\bk-\bq} 
    \label{eq:nafie_RHS_PS}
    \\
    &  +N \frac{\hbar}{m_e} \sum_{\bk',c\neq n} \Im{\frac{\bra{u_{n\bk}}
    \jln{\bm{n}_\bq \cdot(\bhp + \hbar\bk) - \hbar\bq/2}
    \ket{u_{c \bk'}}\bra{u_{c\bk'}}
    \dt{\Gamma}^{A,\bq}
    \ket{u_{n\bk}}}{\epsilon_{n\bk}-\epsilon_{c\bk'}}} \delta_{\bk',\bk+\bq}  \nonumber    
\end{align}
\end{widetext}
where $\dt{\Gamma}$ is the cell-periodic part of $\hbGm$. All details are given in Section IV of the SM .

\begin{figure*}
    \centering    
    \includegraphics[width=\linewidth]{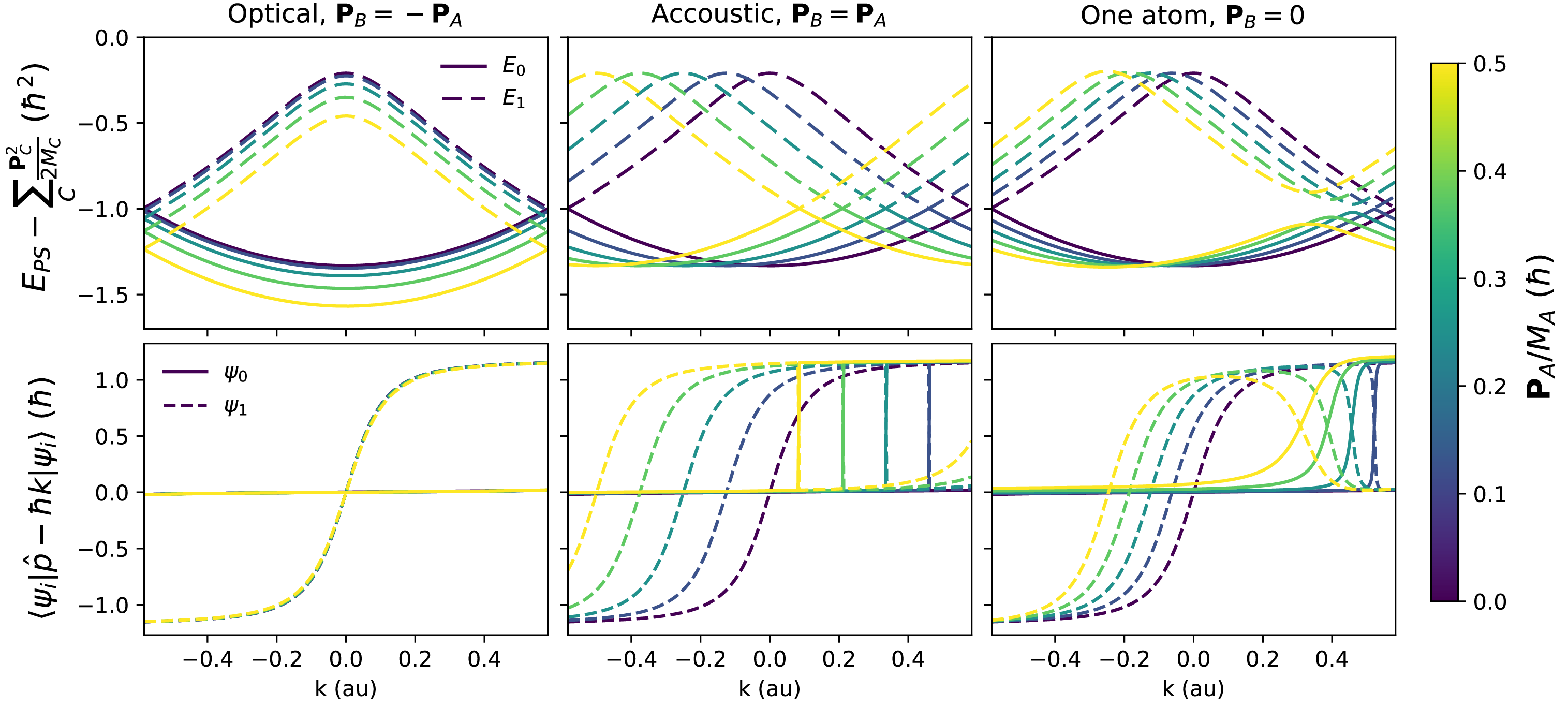}
    \caption{We evaluate both the band energies (top) and the extra electron momentum (bottom)  induced by the nuclear motion for a Hamiltonian consisting of two equivalent  sites in a solid (see SM Section V). Without nuclear momentum, i.e.  $\bP =0.$, the system is a metal (shown in purple).  Here, we plot results for three different scenarios where nuclear motion is included: the presence of an optical phonon $(left)$, an acoustic phonon $(middle)$ and a single atom moving $(right)$.  We plot data for several different magnitudes of the nuclear momentum, represented by the color map on the far right hand side.
    Note that the presence of nuclear motion can lead not only to different electronic energies,  but also to very interesting band structures with new crossings.}
    \label{fig:Pneq0_bands}
\end{figure*}

Our results in Fig. \ref{fig:Nafie_ratio} are  for a model semiconducting potential in 1D (to avoid degenerate bands), as given in SM section V. Indeed we find best agreement between the two methods for $q$ close to zero, with nearly equal quality across all $k$, exactly akin to effective mass theory of solids.  

At the end of the day, the data in Fig. \ref{fig:Nafie_ratio} is a strong indication that, just as in the molecular realm, a phase space electronic structure view of the solid state can indeed restore much of the physics  that is missing when one invokes BO theory and solves the standard electronic structure problem with frozen nuclei.
\bigskip

\paragraph*{The missing electronic momentum in BO based calculations:}

In Fig. \ref{fig:Pneq0_bands} we take one step further and plot the bands $\epsilon^{(PS)}$ and novel electronic momenta, \ul{\em not crystal momenta}, found in the solutions to Eq. \ref{eq:HPS_q0}. This missing momentum is effectively the `anomalous velocity' of the electrons that arises due to the nuclear derivative couplings and which is typically captured by  Berry curvature calculations.\cite{Sundaram1999_NiuEfield,Xiao2010_NiuRMP} For two atoms of equal charge per unit cell, the $\bP=0$ solution for this potential is nearly the same one-dimensional model used in Fig.\ref{fig:Nafie_ratio}, but in this case we take two sites to be symmetric which yields a metallic ground state. In the first column we present the case where the there is an optical phonon of the two atoms. While the $\hat\theta$ function can add distortion to the band energy, there is no symmetry breaking between the two nuclei, and thus no electronic momentum gained in the states. For an acoustic phonon (middle column) with equal nuclear velocities, the effect is similar to translating the entire BZ by the total nuclear momentum, sending $k=0$ to  $k+\bP_{tot} = k+\bP_A+\bP_B$.  The metallic degeneracy here leads to a trivial band crossing, where the character of valence state 0 wavefunction switches abruptly (now resembling the first conduction band) when $\bk+\bP_{tot}$ is larger than the first BZ. In the third column, we show a scenario we translate only one atom and break the symmetry between the nuclei. As in the acoustic picture above, the energetic minima in the BZ shifts by $-\bP_A$, but here we find a finite avoided band crossing (rather than a trivial band crossing). Intuitively, in all cases, for small enough $\bP$, we can model the novel electronic momenta in state valence state 0 and conduction state 1 by interpreting the PS ground state under nuclear motion to be approximately
\begin{align}
&\ket{\psi_{0k}^{PS}(\bP)} \approx \ket{\psi_{0k}(\bP=0)} - 
\\
&\frac{i\hbar}{M_A} \frac{\bra{\psi_{1k}(\bP=0)}\bP_A\cdot\hbGm_A\ket{\psi_{0k}(\bP=0}}{\epsilon_{1k}(\bP=0)-\epsilon_{0k}(\bP=0)} \ket{\psi_{1k}(\bP=0)}, \nonumber 
\end{align}
Indeed, this semiclassical ansatz was the inspiration for Nafie's early work. \cite{Nafie1983}

\paragraph*{Discussion and Conclusions:}
We have introduced a new framework for periodic electronic structure calculations, whereby electronic orbitals and bands are parameterized by the nuclear position $\bR$ and momentum $\bP$.  We submit that the proper, periodic phase space electronic Hamiltonian that must be solved is given in Eq. \ref{eq:HPS_q0}, where a new operator $\bGm$ has been introduced that approximates the derivative coupling in Eq. \ref{eq:dAq} and captures some aspects of  electron-phonon interactions, in particular electronic inertial effects.  Dynamics along an eigensurface of this Hamiltonian are guaranteed to conserve momentum (unlike BO dynamics). To further validate the present theory, we have shown in Fig. \ref{fig:Nafie_ratio} that using $\hbGm$ instead of $d^{A\bq}$, we can recover a BZ-extended Nafie's equality in Eq. \ref{eq:Nafie_equality} with reasonable accuracy. This finding is a strong endorsement that we are capturing the drag that nuclei exert on electrons.
And using the present approach, we have now calculated in Fig. \ref{fig:Pneq0_bands} the band structure as a function of nuclear momentum $\bP$, where one clearly finds that, as q increases, electronic orbitals increases as electrons ``ride the nuclear wave''.

Looking forward,  our theory here has been one-dimensional, but  one important next stage of this research must investigate two and three dimensions, as angular momentum exchange is already known to be critical for many condensed matter experiments, e.g. the Einstein-de Haas effect and the Barnett effect. (Note that Stengel {\em et al} have noted that nuclear motion is especially crucial when dealing with pseudopotentials\cite{Stengel2026}.)  More generally, one would expect the present approach to yield new and non-obvious information about  problems with low-lying electronic states, potentially strong electron-phonon couplings, and multiple minima. For instance, extensions to include spin degrees of freedom and spin-phonon couplings are of immediate interest.
Pushing the equations described above into an {\em ab initio} electronic structure package should be extremely illuminating.

\section*{Acknowledgments}
This work was supported by the U.S. Department of Energy, Office of Science, Office of Basic Energy Sciences, under Award No. DE-SC0025393.

\bibliography{main.bib}
\end{document}


\title{Supplementary Material for: A Phase Space Electronic Structure View of The Solid State} 

\author{Nadine C. Bradbury}
\email{nadinebradbury@princeton.edu}
\author{Linqing Peng}
\author{Joseph E. Subotnik}
\email{subotnik@princeton.edu}
\affiliation{Department of Chemistry, Princeton University, Princeton, NJ USA}
\date{\today}

\maketitle

In this supplementary material (SM), we will both (i) review various background material that is needed to understand the main text of the letter and (ii) fill in gaps in the derivations of the PS electronic structure hamiltonian as presented in the main text. Even though much of the background material may be known to different readers, our instinct is that different readers will be comfortable with different  parts; thus, we have compiled enough background material so that this SM should be self-contained.  As far as filling in gaps from the main text,  we have also aimed to make this SM readable (at the risk of repeating some of the equations in the main text).

{\bf Notation:} We use bold face to indicate indicate vectors in 3 or 3N degrees of freedom. Quantum operators are written with hats.  

\section{Background}
\subsection{Wigner-Weyl Foundations}\label{SM:sec:Wigner-Weyl}
The development of phase space electronic structure theory is intimately tied to Wigner-Weyl transforms and their symbols.
A very important fact is that Wigner-Weyl transforms commute with linear canonical transformations $K$ with determinant one--this holds for real or complex-valued $K$ transformations. For instance,  consider a  system described by canonical coordinates $(\bx,\bp)$ and which are transformed  to $(\br,\bpi)$:

\begin{align}
\label{eq:xr}
    \br &= K \bx \\
    \label{eq:ppi}
    \bpi &= (K^{-1})^T \bp
\end{align}

\subsubsection{Path 1:Wigner Transform then Change Coordinates}
The Wigner transform of $\hO$ is:
\begin{align}
    O_W(\bx,\bp) = \int d\bDelta  \left<x  + \frac{\bDelta}{2} \middle | \hO \middle| x  - \frac{\bDelta}{2} \right> e^{i \bp \cdot \bDelta }
\end{align}
Thereafter, if switch to $(\br,\bpi)$ coordinates, we find:
\begin{align}
     \tilde{O}_W\left(\br,\bpi \right)
    &= O_W\left(K^{-1} \br, K^T \bp \right) \\
    & = \int d\bDelta 
    \left<K^{-1} \br  + \frac{\bDelta}{2} \middle | \hO \middle| K^{-1} \br  - \frac{\bDelta}{2} \right> e^{i (K^T\bpi) \cdot \bDelta }
    \label{eq:check1gen}
\end{align}

\subsubsection{Path 2: Change Coordinates then Wigner Transform}
The alternative is to change coordinates first:
\begin{align}
\left<\br \middle | \tilde{\hO} \middle| \br'\right>  = 
\left<K^{-1}\br \middle | \hO \middle| K^{-1}\br' \right> \frac{1}{|\det(K^{-1})|}     
\end{align}

The determinant factor arises whenever one changes coordinates as seen most easily by insisting that a wavefunction still be normalized.
The Wigner transform becomes:
\begin{align}
\tilde{O}_W\left( \br,\bpi \right)
&=\int d\bDelta' \left<\br  + \frac{\bDelta'}{2} \middle | \tilde{\hO} \middle| \br - \frac{\bDelta'}{2}\right> e^{i \bpi \cdot  \bDelta'}
\frac{1}{|\det(K^{-1})|}  
\\
& = \int d\bDelta'  \left<K^{-1} (\br  + \frac{\bDelta'}{2})\middle | \hO \middle| K^{-1} (\br  - \frac{\bDelta'}{2})\right>  e^{i \bpi \cdot \bDelta'}
\frac{1}{|\det(K^{-1})|} 
\label{eq:check2a}
\end{align}

Now, we change integration variables:
\begin{align}
    \bDelta' & = K \bDelta
\end{align}

\begin{align}
\tilde{O}_W\left( \br,\bpi \right)
& = \frac{|\det(K^{-1})|}{|\det(K^{-1})|}  \int d\bDelta  \left<K^{-1} \br  + \frac{\bDelta}{2}\middle | \hO \middle| K^{-1} \br  - \frac{\bDelta}{2}\right>  e^{i \bpi \cdot (K \bDelta)}
\label{eq:check2}
\end{align}

  Thus,  if one performs a  Wigner transformation from operators to symbols, followed by an  arbitrary linear changes of coordinates, and then one performs one final Weyl transformation from symbols back to operators, one will find that quantum system is entirely unchanged.  
  Note that the  argument  above holds if $K$ is real-valued or complex-valued.  
  
  Note also that the equality
  \begin{align}
    \left(\hat{A}\hat{B}\right)_W = \hat{A}_W \star \hat{B}_W 
  \end{align}
  holds in the $(\bx,\bp)$ or $(\br,\bpi)$ coordinates or any other linearly transformed coordinates (real or complex-valued). This  fact is easily proven because $\star = \exp(\overleftrightarrow{\Lambda}_{\bx,\bp})$
  where $\overleftrightarrow{\Lambda}_{\bx,\bp}= \frac{i\hbar}{2} \left(\overleftarrow{\partial_\bx}\overrightarrow{\partial_\bp} - \overleftarrow{\partial_\bp}\overrightarrow{\partial_\bx}\right)$
  and for any  linear transformation as in Eqs. \ref{eq:xr}-\ref{eq:ppi}, it follows that  $\overleftrightarrow{\Lambda}_{\bx,\bp} = \overleftrightarrow{\Lambda}_{\br,\bpi}$.

\subsection{BO band structure theory as viewed through a Wigner Transformation }

Here, we review standard band stucture within the Born-Oppenheimer approximation.
We begin by treating  electrons with second quantization, and we treat the nuclei with first quantization. 
\begin{align}
\hH &= \frac{\bhp^2}{2m_e} + V(\bhr,\{\hbR_{A j}\}) + \sum_{Aj} \frac{\hbP^2_{Aj}}{2M_{Aj}} \label{eq:H_periodic_full quantum}
\end{align}
Here, $A= 1,2,\ldots,N_A$ where
$N_A$ is the number of atoms in one unit cell and $j = 1,2, \ldots, N$ where $N$ is the number of  unit cells. So there are totally $N_A N$ atoms. Let the standard unit cell have length $a$ so the total periodic solid has length $N a$.

The very first thing we do is take the Wigner transform over the nuclei:

\begin{align}
\hH_W &= \frac{\bhp^2}{2m_e} + V(\bhr,\{\bR_{A j}\}) + \sum_{Aj} \frac{\bP^2_{Aj}}{2M_{Aj}}
\end{align}

Now, the  standard band theory problem is to diagonalize the  above electronic Hamiltonian:
\begin{align}
    \hH_W(\bR,\bP) \ket{\psi_j} = \epsilon_j \ket{\psi_j}
\end{align}
using $\bR$ and $\bP$ as shorthand for the full set of $\{\bR_{Aj}, \bP_{Aj}\}$.

Now, because of the exact periodicity over length $Na$, $\ket{\psi}$ must satisfy:
\begin{align}
    \psi_j(\br+N\ba_i) = \psi_j(\br)
\end{align}
for $i = 1,2,3$ and $\ba_i$ being a primitive-cell lattice vector.

More interestingly, if the nuclei are arranged exactly periodically over a single unit cell (say, at positions $\bR_0$), then the diagonalization simplifies and we can write (Bloch):
\begin{align}
\ket{\psi_j} &= \ket{\psi_{n \bk}} \\
\psi_{n \bk}(\br+\ba_i,\bR) &= e^{i\bk\cdot\ba_i} \psi_{n \bk}(\br,\bR) 
\label{eq:bloch0}
\end{align}
where $\bk = j_1 \bm{b}_1 + j_2 \bm{b}_2 + j_3\bm{b}_3$ for some triple of integers $j$ and reciprocal lattice vectors $\left\{ \bm{b}_j \right\}$ defined such that $\ba_i \cdot \bm{b}_j = 2\pi \delta_{ij}$ for primitive cell lattice vectors $\left\{ \ba_i \right\}$ .

In practice, the symmetry above implies that we can write:
\begin{align}
    \psi_{n\bk}(\br,\bR) &= e^{i\bk\cdot\br} u_{n\bk}(\br,\bR) 
\label{eq:bloch}
\end{align}
Here, $u_{\bk}$ is  a solution to:
\begin{align}
\hat{H}^W_{\bk} u_{n\bk}(r) = 
\left( \frac{(\bhp + \hbar \bk)^2}{2m_e} + V(\bhr,\{\bR_{A j}\}) + \sum_{Aj} \frac{\bP^2_{Aj}}{2M_{A}} \right) u_{n\bk}(\br) = \epsilon_{n\bk} u_{n\bk}(\br)
\label{eq:firstHk}
\end{align}
and is periodic over the unit cell:
\begin{align}
 u_{n\bk}(\br \pm \ba_i) = u_{n\bk}(\br) 
\end{align}
Note that, in Eq. \ref{eq:firstHk}, we have defined:
\begin{eqnarray}
    \hat{H}^W_{\bk} = e^{-i\bk\cdot \bhr} \hH_W e^{i\bk\cdot\bhr} 
\end{eqnarray}
as is standard nomenclature.

The bottom line is that, when the atoms are exactly periodic over each unit cell, one can write:
\begin{align}
    \hH_W = \sum_{n\bk} \ket{\psi_{n\bk}(\bR)}\epsilon_{n\bk}(\bR,\bP)\bra{\psi_{n\bk}(\bR)}
 \label{eq:key}
\end{align}

Note that, for the diagonals in Eq. \ref{eq:key}, the energy has two components, the potential energy ${\epsilon}^{(0)}(\bR)$ and the kinetic energy ${\epsilon}^{(1)}(\bP)$:

\begin{align}
{\epsilon}(\bR,\bP) &= 
{\epsilon}^{(0)}(\bR) + 
{\epsilon}^{(1)}(\bP) \\
{\epsilon}^{(1)}(\bP) &=
\sum_{A\ell} \frac{\bP_{A\ell} \cdot \bP_{A\ell}}{2M_A} 
\end{align}

\section{Review of The Electron-Phonon Hamiltonian as viewed from a BO perspective with a Wigner Transformation}

It is crucial to emphasize that, even without exact periodicity over one unit cell, the representation of the Wignerized Hamiltonian in Eq. \ref{eq:key} is still meaningful. 
After all, one can always diagonalize $\hH_W$, generate eigenvectors/values, and  then match them up continuously, i.e. analytically continuing $\psi_{n\bk}(\br,\bR)$ from the perfectly periodic regime to the slightly disordered regime. The only difference is that Bloch's theorem (Eq. \ref{eq:bloch0}) no longer holds:

\begin{align}
\psi_{n\bk}(\br\pm \ba_i) &\ne e^{\pm i \bk \cdot \ba_i}\psi_{n\bk}(\br) 
\end{align}

That being said, the decomposition in  Eq. \ref{eq:bloch} is still a valid ansatz as the $u_{n \bk}$ can be extended by continuity.  
However, the difference is that now the $u_{nk}(\br)$ functions are not periodic over the unit cell:
\begin{align}
 u_{n\bk}(\br \pm \ba_i) &\ne u_{n\bk}(\br) 
\end{align}
In the end, Eq. \ref{eq:key} still holds (with a slightly different meaning for $\bk$).

\subsection{A Fourier Transform Over The Nuclei
}

At this point, one would like to switch coordinates for the nuclei and apply a fourier transform, changing from $\bR_{\ell}$ to $\bQ_\bq$ and from 
$\bP_\ell$ to $\bPi_\bq$:
\begin{align}
    \bQ_{A\bq} &=  \frac{1}{\sqrt{N}}\sum_{\bR_{\ell}} e^{- i \bR_{\ell}\cdot \bq} (\bR_{A\ell} - \bR_{A\ell}^0) \\
    (\bR_{A\ell} -\bR_{A\ell}^0) &=  \frac{1}{\sqrt{N}}\sum_{\bq} e^{i \bR_{\ell}\cdot \bq } \bQ_{A\bq}  \\
    \bPi_{A\bq} &= \frac{1}{\sqrt{N}}\sum_{\bR_{\ell}} e^{i \bR_{\ell}\cdot\bq}\bP_{A\ell} \\
    \bP_{A\ell} &= \frac{1}{\sqrt{N}} \sum_\bq e^{-i \bR_{\ell}\cdot\bq} \bPi_{A\bq}
\end{align}
where $\bq$ is defined similarly to $\bk$ for the triple of integers, and $\bR_\ell = \bm{\ell} \cdot \ba =  \ell_1 \ba_1 + \ell_2\ba_2+\ell_3\ba_3$ for the triple of integers $\bl$.
Note that the quantized versions of $\bQ_{A\bq}$ and $\bPi_{A\bq}$ are not hermitian operators; instead, e.g. $\hat{\bQ}^{\dagger}_{A\bq} = \hat{\bQ}_{A-\bq}$  

To make progress, we must now also write down the electronic eigenfunctions $\psi_{n \bk}$ and eigenvalues $\epsilon_{n \bk}$ relative to these new nuclear coordinates:

\begin{align}
    \psi_{n\bk}(\br,\bQ) &\equiv
    \psi_{n\bk}(\br, \bR(\bQ))  \\
    \epsilon_{n\bk}(\bQ,\bPi) &= \epsilon_{n\bk}( \bR(\bQ),\bP(\bPi))
\end{align}
Henceforth, unless specified otherwise, $\psi$ and $\epsilon$ are presumed to be functions ($\bQ$,$\bPi$) rather than $(\bR,\bP)$. In these new coordinates, the Wignerized electronic Hamiltonian then takes the form:
\begin{align}
    \hH_W(\bQ,\bPi) = \sum_{n\bk} \ket{\psi_{n\bk}(\bQ)}\epsilon_{nk}(\bQ,\bPi)\bra{\psi_{n\bk}(\bQ)}
\end{align}

The BO approximation is equivalent to a Weyl transform of the diagonal eigenvalue matrix:

\begin{align}
    \left(\hat{H}^{BO}\right)_{n \bk, n\bk'} &= \delta_{\bk,\bk'} W^{-1}(\epsilon_{n\bk}(\bQ,\bPi)) \label{eq:hH_BO}
\end{align}
and the diagonal elements still decompose as potential plus kinetic energies:
\begin{align}
\epsilon(\bQ,\bPi) &= 
\epsilon^{(0)}(\bQ) + 
\epsilon^{(1)}(\bPi) \\
\epsilon^{(1)}(\bPi) &=
\sum_{A \bq} \frac{\bPi_{A\bq} \cdot \bPi_{A -\bq}}{2M_A} 
\end{align}

\subsection{The Standard Born-Huang Expansion}

To go beyond the BO approximation and recover the true Hamiltonian, we must compute the Weyl transform of:
\begin{align}
        \hH = W^{-1}(\hH_W(\bQ,\bPi)) = W^{-1} \left( \sum_{n\bk} \ket{\psi_{n\bk}(\bQ)}\epsilon_{n\bk}(\bQ,\bPi)\bra{\psi_{n\bk}(\bQ)} \right)
\end{align}

Unfortunately, the process above is difficult and requires a complete basis. We want to work with only a limited number of electronic states, and so we want to project the Hamiltonian down. To that end,  imagine we have  a basis, $\left\{ e_y \right\}$ in which we expand the eigenvectors:
\begin{align}
    Z_{y,nk}(\bQ) \equiv \left< y \middle | \psi_{n\bk} \right>
\end{align}
Because $Z$ is a unitary matrix and because similarity transformations preserve an eigenspectrum, let us construct a  Wignerized final electron-phonon Hamiltonian of the following form: 
\begin{eqnarray}
    \hat{H}'_W = Z^{\dagger}(\bQ) * \hH_W(\bQ,\bPi) * Z(\bQ)  \label{eq:first_star}
\end{eqnarray}
A Weyl transform of $\hat{H}'_W$ carries all of the information of the original hamiltonian insofar as 
\begin{align}
    W^{-1}\left(\hat{H}'_W\right)  = \hat{Z}^{\dagger} \hat{H} \hat{Z}
\end{align}
using the rules of the Moyal star product.

Finally, let us evaluate Eq. \ref{eq:first_star}.  If we define quantities,
\begin{align}
\label{eq:def_d}
    \hat{\bd}^{A \bq}  &\equiv Z^{\dagger} \frac{\partial Z}{\partial \bQ_{A \bq}}\\
    \label{eq:def_zeta}
     \zeta &\equiv \sum_{Aq} \frac{\jln{\hbar^2}}{2M_A} \frac{\partial Z^{\dagger}}{\partial \bQ_{A -\bq}} \frac{\partial Z}{\partial \bQ_{A \bq}} \\
\end{align}
one can show easily that Eq. \ref{eq:first_star} becomes:
\begin{align}
    \hat{H}'_W = \epsilon^{(0)}(\bQ) + \sum_{A \bq} \frac{\bPi_{A \bq} \cdot \bPi_{A -\bq}}{2M_A}   - i \hbar \sum_{A \bq} \frac{\bPi^{A\bq} \cdot \hat{\bd}^{A-\bq}}{M_A} + \hat{\zeta}\label{eq:SM:H_QShenvi}
\end{align}

If we want to write this expression out in  electronic index form, the result is

\begin{align}
    (\hat{H}'_W)_{m\bk',n\bk} =  \delta_{mn}\delta_{\bk,\bk'} \left( \epsilon^{(0)}(\bQ) +   \sum_{A \bq} \frac{\bPi_{A \bq} \cdot \bPi_{A -\bq}}{2M_A} \right)  - i \hbar \sum_{A \bq} \frac{\bPi^{A \bq} \cdot \bd^{A -\bq}_{mk',nk}}{M_A} + \zeta_{mk',nk} \label{eq:H_QShenvi_index}
\end{align}
where

\begin{align}
\bd^{A\bq}_{m\bk',n\bk} & = \left<\psi_{m\bk'} \middle| \frac{\partial }{\partial \bQ_{A\bq}} \psi_{n\bk}\right> 
\label{eq:omg2}\\
\zeta_{m\bk',n\bk}  &= \sum_{A \bq} \jln{\frac{\hbar^2}{2M_A}} \left< \frac{\partial }{\partial \bQ_{A\bq}} \psi_{m\bk'} \middle| \frac{\partial }{\partial \bQ_{A,\bq}} \psi_{n\bk}\right> 
\label{eq:omg1}
\end{align}

\subsection{The First Order Response In The Limit of a Periodic Lattice}
Eq. \ref{eq:H_QShenvi_index} is the Born-Huang expansion for a solid-state problem. That being said, it remains to evaluate the derivative coupling $\hat{\bd}^A$ in Eq. \ref{eq:def_d} ($\zeta$ is usually dropped).
Calculating ${\hat{\bd}}^A$ is an arduous task that is equivalent to evaluating the wavefunction response    $\partial \ket{\psi_{n\bk}(\bQ)}/\partial \bQ_{A\bq}$, but one that is quite feasible at a geometry $\bR_0$ that is periodic over each and every unit cell. 

Imagine that, starting from a unit cell-periodic $\bR_0$ configuration, we   slightly displace the nuclei with a perturbation of wavelength $\bq$, so that $\bQ_\bq \ne 0$. The resulting potential will be of the form: 
\begin{align}
    \hat{V}_{new} &= \hat{V}_0 + \delta \hat{V}\\
    \delta \hat{V}(\br) &=  \sum_{A\bR_\ell} \frac{\partial \hat{V}(\br)}{\partial \bR_{A\ell}} (\bR_{A\ell}-\bR_{A\ell}^0) \\
    & =   \sum_{\bq\bR_l A} \frac{\partial \hat{V}(\br)}{\partial \bR_{A\ell}} e^{i\bR_l \cdot \bq} Q_{A\bq} \\
    & =   \sum_{\bq\bR_l A} \frac{\partial \hat{V}(\br)}{\partial \bR_{A\ell}} e^{-i(\br-\bR_l)\cdot \bq} e^{i\br\cdot\bq} \bQ_{A\bq} 
    \end{align}
    
Now, note that (up to a kinetic energy) $V(\br) =\sum_{A\ell} V_{A\ell}(\br)$ and $V_{A\ell}(\br)$ is a function of $\br-
\bR_{A\ell}$ so that, if we define
\begin{align}
    \bR_{\ell'} = (\ell_1-1)\ba_1 + \ell_2\ba_2 +\ell_3\ba_3
\end{align}
then
\begin{align}
    V_{A,\ell}(\br+\ba_1) &= V_{A,\ell'}(\br) \\
     \frac{\partial V_{A,\ell}(\br+\ba_1)}{\partial \bR_{A\ell}} &= 
    \frac{\partial V_{A,\ell'}(\br)}{\partial \bR_{A,\ell'}}
\end{align}
The result is similar for shifts by $\ba_2$ and $\ba_3$.
Thus, the function
\begin{align}
    \delta \hat{W}^{A\bq}(\br) &\equiv \sum_{\bR_l} \frac{\partial \hat{V}_{A\ell}(\br)}{\partial \bR_{A\ell}} e^{-i(\br-\bR_{\ell})\cdot\bq} 
\end{align}
is periodic, satisfying
\begin{align}
    \delta \hat{W}^{A\bq}(\br+\ba_1) & = \sum_{\bR_l} \frac{\partial \hat{V}_{A\ell}(\br+\ba_1)}{\partial \bR_{A\ell}} e^{-i(\br+\ba_1-\bR_{\ell})\cdot\bq} \\
    & = \sum_{\bR_{\ell'}} \frac{\partial \hat{V}_{A\ell'}(\br)}{\partial \bR_{Al'}} e^{-i(\br-\bR_{\ell'})\cdot\bq} \\
    & = \delta \hat{W}^{A\bq}(\br)
\end{align}
Here, we have used a dummy variable $\ell'$ as defined above for $\bR_{\ell'}$.
Note that $\delta \hat{W}^{A \bq}$ is not hermitian, but satisfies 

\begin{align}
\delta \hat{W}^{A,-\bq}
    = \left(\delta \hat{W}^{A,\bq} \right)^{\dagger}
\end{align}

In the end, it follows that:
    \begin{align}
    \delta \hat{V} & = \sum_{A\bq} Q_{A\bq} e^{i\bq\cdot \br} \delta \hat{W}^{A \bq} (\br) 
    \label{eq:deltaVq}
\end{align}
where $\delta \hat{V}$ {\em is} hermitian (which follows since $Q_{A -\bq} = Q_{A \bq}^*$).

We can now differentiate $\ket{\psi_{nk}(\bQ)}$.  Starting at $\bQ=0$ (the configuration periodic over the unit cell), first order perturbation theory dictates that:

\begin{align}
\delta \ket{\psi_{nk}(\bQ)} &=
    \sum_{pk'} \ket{\psi_{p\bk'} (\bQ = 0)} \frac{\left< \psi_{p\bk'}(\bQ = 0) \middle| \delta \hat{V}\middle | \psi_{n\bk} (\bQ = 0) \right>}{\jln{\epsilon_{n\bk}-\epsilon_{p\bk'} } }\\
    &=
    \sum_{p\bk'A\bq} \ket{\psi_{p\bk'} (\bQ = 0)} \frac{\left< \psi_{p\bk'}(\bQ = 0) \middle| Q_{A\bq} e^{i\bq\cdot\br} \delta \hat{W}^{A\bq}(\br) \middle | \psi_{n\bk} (\bQ = 0) \right>}{\jln{\epsilon_{n\bk}-\epsilon_{p\bk'} }}
\end{align}

In other words, for a small displacement in $\bQ_{A\bq}$, we find:
\begin{align}
    \frac{\partial \ket{\psi_{n\bk}(\bQ)}}{\partial \bQ_{A\bq}}\bigg|_{\bQ = 0} =
    \sum_{p\ne n,\bk'} \ket{\psi_{p\bk'} (\bQ = 0)} \frac{\left< \psi_{p\bk'}(\bQ = 0) \middle| \delta \hat{W}^{A \bq} e^{i\bq\cdot\br}\middle | \psi_{n\bk} (\bQ = 0) \right>}{\jln{\epsilon_{n\bk}-\epsilon_{p\bk'} }}\label{eq:forbelow}
\end{align}

Importantly, here we have used intermediate normalization above so that $pk'$ cannot be the same as $nk$. Furthermore, because occupied states cannot be doubly occupied, if $n$ is occupied, we will assume that $p$ must be in the  conduction band (henceforward labeled
$c$). 

Now, using Eq.  \ref{eq:bloch},
it follows that
\begin{align}
\left< \psi_{p\bk'}(\bQ = 0) \middle| \delta \hat{W}^{A \bq} e^{i\bq\cdot \br} \middle | \psi_{n\bk} (\bQ  = 0) \right> 
&= \int_0^{N\ba} d\br \; u^*_{p\bk'}(r,\bQ = 0)
e^{i(\bk-\bk'+\bq)\cdot\br}\delta \hat{W}^\bq_A(\br)
u_{n\bk}(\br,\bQ = 0)\\
&=  N \delta_{\bk',\bk+\bq} \int_0^{\ba} d\br \; u^*_{p\bk'}(\br,\bQ = 0)
\delta \hat{W}^{A \bq}(r)
u_{n\bk}(\br,\bQ = 0) 
\label{eq:define_end1}
\end{align}

As should be clear from the above manipulations, one must be very careful to distinguish between the $\psi_{n \bk}$ and $u_{n \bk}$
wavefunctions, as they have different matrix elements.
In this supplementary information, all operators are  taken with respect to the $\psi_{n\bk}$ basis unless stated otherwise. 
That being said, if we make the definition
\begin{align}
    \left< u_{p\bk'} \middle|\delta \hat{W}^{A\bq} \middle| u_{n\bk} \right> = \int_0^{\ba} d\br \; u^*_{p\bk'}(\br,\bQ = 0)
\delta \hat{W}^{A \bq}(\br)
u_{n\bk}(\br,\bQ = 0), 
\label{eq:afterk1}
\end{align}
we can also write Eq. \ref{eq:define_end1} above as 
\begin{align}
         \left< \psi_{p\bk'}(\bQ = 0) \middle| \delta \hat{W}^{A\bq} e^{i\bq\cdot \br} \middle | \psi_{n\bk} (\bQ  = 0) \right>
         &\equiv N  \left< u_{p\bk'} \middle|\delta \hat{W}^{A\bq} \middle| u_{n\bk} \right> \delta_{\bk',\bk+\bq}
         \label{eq:afterk2}
\end{align}

Finally, putting it all together, we find
\begin{align}
    \frac{\partial \ket{\psi_{n\bk}(\bQ)}}{\partial \bQ_{A\bq}}\bigg|_{\bQ = 0} =N
 \sum_{p \ne n} \ket{\psi_{p\bk+\bq} (\bQ = 0)} \frac{\left< u_{p, \bk+\bq}  \middle| \delta \hat{W}^{A\bq} \middle | u_{n,\bk} \right>}{\epsilon_{p,\bk+\bq} - \epsilon_{n,\bk}} \label{eq:dQpsi}
\end{align}
or equivalently in short-hand notation:
\begin{align}
    \left<\psi_{c\bk'} \middle| \partial_{\bQ_{A\bq}}\psi_{n\bk} \right> = N  \delta_{\bk',\bk+\bq}
\frac{\left< u_{c, \bk+\bq}  \middle| \delta \hat{W}^{A\bq} \middle | u_{n,\bk} \right>}{\epsilon_{c,\bk+\bq} - \epsilon_{n,\bk}} \label{eq:dQpsi2}
\end{align}

The final expressions for the quantities in Eqs. \ref{eq:omg2}-\ref{eq:omg1} then become:

\begin{align}
     \bd^{A\bq}_{n\bk',m\bk} &= 
        \frac{\bra{\psi_{n\bk'}} \delta \hat{W}^{A\bq} e^{i \bq \cdot \br}  \ket{\psi_{m\bk}}}{\epsilon_{n\bk'}-\epsilon_{m\bk}} \label{eq:d^A_q}
     \\
     &= N\frac{\bra{u_{n\bk'}} \delta \hat{W}^{A\bq}  \ket{u_{m\bk}}}{\epsilon_{n\bk+\bq}-\epsilon_{m\bk}}\delta_{\bk',\bk+\bq} \label{eq:dAq}
\end{align}
     and
\begin{align}
     \zeta_{n\bk',m\bk} &= N \sum_{A\bq p} \frac{1}{2M_A} 
     \frac{  \bra{u_{n \bk'}} \delta \hat{W}^{A,-\bq}  \ket{u_{p \bk+\bq}}  \bra{u_{p \bk+\bq}} \delta \hat{W}^{A\bq}  \ket{u_{m \bk}}}{(\epsilon_{n,\bk'}-\epsilon_{p,\bk+\bq}) (\epsilon_{p,\bk+\bq}-\epsilon_{m,\bk}) } \delta_{\bk',\bk}
     \label{eq:zeta}
\end{align}

Note that 
\begin{eqnarray}
\left(d_{nk',mk}^{A
\bq}\right)^* =- d_{mk,nk'}^{A-\bq}
\end{eqnarray}
or in more abstract form
\begin{eqnarray}
\left(d^{A
\bq}\right)^{\dagger} =- d^{A-\bq}
\end{eqnarray}
The total Hamiltonian in Eq. \ref{eq:H_QShenvi_index} is hermitian (given that $\bPi_{A -\bq} = \bPi_{A \bq}^*$).

\subsection{Symmetries of the Electron-Phonon Matrix Elements}

It is very easy to show that:
\begin{align}
    -i\hbar\sum_A d^{A\bq=0}_{n\bk',m\bk} &=\left< \psi_{n
    \bk'} \middle| \bhp \middle| \psi_{m\bk} \right>
    \label{eq:sumdA0}
\end{align}
This symmetry arises due to translational invariance of the potential
\begin{align}
    \sum_A \delta \hat{W}^{A\bq=0}(\br) + \frac{\partial \hat V}{\partial \br} = 0
\end{align}
Thus,
\begin{align}
    -i\hbar \sum_A \delta \hat{W}^{A\bq=0}(\br) =  \left[ \hat V,  \bhp \right] =
     \left[ \hat H_{el},  \bhp \right]
\end{align}
which can be plugged into Eq. \ref{eq:d^A_q}.

Second, due to translational invariance of the entire Hamiltonian in Eq. \ref{eq:H_QShenvi_index},
\begin{align}
    \sum_B \frac{\partial}{\partial \bQ_{B\bq=0}}\bd^{A\bq}_{n\bk,m\bk'} = 0 \label{eq:nablaBdAq}
\end{align}
The symmetry in Eq. \ref{eq:d^A_q} reflects the fact that the eigenvalues of $\hH_{el}$ and the matrix elements $\bra{u_{m\bk}}\delta \hat{W}\ket{u_{n\bk}}$ depend only on the relative positions of the nuclei (and not their absolute positions in space).

\subsection{Connection to Born-Huang Phonons and Traditional Phonon Theory }
The combination of Shenvi's Hamiltonian as given in Eq. \ref{eq:SM:H_QShenvi}, the $\bq$-dependent derivative coupling and second order term given in Eqs. \ref{eq:d^A_q} and \ref{eq:zeta}, and a Weyl transform over $\bQ_{A\bq}$ yields a vibrational Hamiltonian that contains full electron-phonon couplings as prescribed by  the standard Born-Huang vibrational Hamiltonian. For a fixed $\bq$, the Hamiltonian reads:
\begin{align}
    \hat{H}^{BH} = \epsilon^{(0)}(\bhQ_{\bq}) + \sum_{A\bq} \frac{\bhPi_{A\bq }\cdot \bhPi_{A,-\bq}}{2M_A}   
    - i \hbar  \sum_{A\bq}  \frac{\bhPi_{A\bq} \cdot \hat{\bd}^{A,-\bq}}{M_A} 
    + \hat{\zeta}
\label{eq:H_QBornHuang}
\end{align}

Traditional phonon theory (and  density functional perturbation theory more generally) is usually framed as a``monochromatic'' (meaning one $q$ at a time) perturbation theory where one starts with the BO vibrational Hamiltonian given in Eq. \ref{eq:hH_BO}-- i.e. one drops the $\hat{\bd}^{A\bq}$ and $\hat{\zeta}$ terms, and removes the sums over $\bq$ from Eq. \ref{eq:H_QBornHuang} above such that each $\bq$ is solved independently. The  potential energy surface $\epsilon^{(0)}_{n\bk}(\bQ)$ is then harmonically expanded around $\bQ=0$.
\begin{align}
    \left(\hH_{BO}\right)^{\bq}_{n \bk} &\approx  \sum_{AB} \left[ \frac{\hbPi_{A,-\bq}\cdot \hbPi_{A\bq}}{2M_A}\delta_{AB} +   \hat{\bQ}_{A,-\bq} \cdot \left.\frac{\partial^2 \epsilon^{(0)}_{n\bk}}{\partial \bQ_{A,-\bq}\partial \bQ_{B\bq}}\right|_{\bQ=0} \cdot \hbQ_{B,\bq} \right]
    \label{eq:mc}
\end{align}
where the hessian takes the form:
\begin{align}
      \frac{\partial^2 \epsilon^{(0)}_{n\bk}}{\partial \bQ_{A,-\bq}\partial \bQ_{B\bq}}
      = \left[ \left\langle\psi_{n\bk}\middle|\frac{\partial^2 H^W}{\partial \bQ_{A,-\bq}\partial \bQ_{B\bq}}\middle|\psi_{n\bk}\right\rangle + \left\langle\frac{\partial\psi_{n\bk}}{\partial \bQ_{A\bq}}\middle|\frac{\partial H^W}{\partial \bQ_{B\bq}}\psi_{n\bk}\right\rangle +  \left\langle  \psi_{n\bk} \middle|
    \frac{\partial H^{W}}{\partial \bQ_{B\bq} } \frac{\partial\psi_{n\bk}}{\partial \bQ_{A,-\bq}}\right\rangle\right]\Bigg|_{\bQ=0} 
\end{align}
which reduces over the unit cell to:
\begin{align}
     & \frac{1}{N} \frac{\partial^2 \epsilon^{(0)}_{n\bk}}{\partial \bQ_{A,-\bq}\partial \bQ_{B\bq}}
    = \left\langle u_{n\bk}\middle|\frac{\partial^2 H_{\bk}^W}{\partial \bQ_{A,-\bq}\partial \bQ_{B\bq}}\middle| u_{n\bk}\right\rangle \\
    & + \sum_{\bk',p\neq n} \Biggl( 
    \frac{\delta_{\bk',\bk+\bq}}{\epsilon_{p\bk'}-\epsilon_{n\bk}}
    \bra{u_{n\bk}}\delta \hat{W}^{A,-\bq}\ket{u_{p\bk'}}
\bra{u_{p\bk'}}\delta \hat{W}^{B\bq} \ket{u_{n\bk}} + \bra{u_{n\bk}}\delta \hat{W}^{B\bq}\ket{u_{p\bk'}}
\bra{u_{p\bk'}}\delta \hat{W}^{A, -\bq} \ket{u_{n\bk}}
\nonumber
    \Biggr) 
\end{align}
Up to mass weighting, the so-called `dynamical' matrix 
\begin{align}
    D^{\bq}_{AB} \equiv \frac{\partial^2 \epsilon^{(0)}_{n\bk}}{\partial \bQ_{A,-\bq}\partial \bQ_{B\bq}} 
\end{align}
is hermitian insofar as $\left(D^{\bq}_{AB}\right)^* = D^{\bq}_{BA}$
with an extra symmetry in that
$\left(D^{\bq}_{AB}\right)^* = D^{-\bq}_{AB}$ from real space translational invariance. 
Phonons arise by diagonalizing the Hamiltonian in Eq. \ref{eq:mc}; for a system with many electrons, of course we diagonalize $\sum_{n\bk} \left(\hH_{BO}\right)^{\bq}_{n \bk}.$

The formalism above is akin to a wavefunction form of standard density functional perturbation theory \cite{Baroni2001_DFPT}, though the notation differs slightly due to the difference in derivation.
Note that \jln{$\left(D^{\bq}_{AB}\right) \ne D^{-\bq}_{AB}$} so that formally one can still recover different phonon eigenfunctions for $\bq$ vs $-\bq$; the eigen-energies will be equal for a standard, time-reversible invariant Hamiltonian but can be different if one includes the $\bd \cdot \bPi$ terms from Eq. \ref{eq:H_QBornHuang} that break time-reversal symmetry when \jln{$\bq$ couples to spin}. Under the harmonic hessian approximation these time-reversal breaking terms have been studied in the context of both derivative couplings and external magnetic fields, for both in model and \emph{ab-initio} contexts.\cite{Qin2012,Komiyama2021,Bonini2023,Tellgren2023}

\subsection{How To Understand Momentum Conservation}

Linear momentum conservation holds both for systems in free space and for systems with periodic boundary conditions.\cite{frenkel_and_smit}  Momentum conservation within BO theory arises by ignoring the electronic dynamics and  writing the total momentum as:
\begin{align}
\bP_{tot} = \sum_{A\bR_l} \bP_{A\ell}    
\end{align}

Thereafter, if we are moving along surface $nk$, note that:
\begin{align}
    \sum_A \dot{\bP}_{A\ell} = -\sum_{A\bR_l} \frac{\partial \epsilon_{n\bk}}{\partial \bR_{A\ell}} = - \sum_A \frac{\partial \epsilon_{n\bk}}{\partial \bQ_{A0}}  
    = 0 
\end{align}

\section{Nafie's equality Within a BO framework and a Wigner Transformation}

In the main body of the text, one of our central results is a proof of Nafie's theorem for periodic systems. Before offering up the proof below, let us review the standard result for motion along a molecular BO state without any periodicity.

\subsection{The Nafie Equality for Molecules (i.e. without Periodicity)}\label{sec:nafie_mol}
As a reminder to the reader, Nafie's celebrated dipole equality in molecules takes the form
\begin{equation}
\label{eq:nafie_molecular}\left.\frac{\partial\langle \br \rangle}{\partial \bR_A}\right|_{\bP = 0} = \frac{M_A}{m_e}\left.\frac{\partial\langle \bp  \rangle}{\partial \bP_A}\right|_{\bP = 0},
\end{equation}
where we have introduced  factors of mass; Nafie's  original studied worked with  \emph{velocities} so these mass factors did not appear.\cite{Nafie1983} Nonetheless, the proof proceeds the same (as follows).

Consider a BO electronic wavefunction $\ket{\Phi_n}$ that satisfies $\bra{\Phi_n}\bhp\ket{\Phi_n} = 0$. (Here, we will use capital $\Psi$ and $\Phi_n$ to distinguish molecular states from Bloch states elsewhere in this paper.)  If one uses perturbation theory to correct for the effect of nuclear motion, the result is simple:
\begin{align}
    \ket{\Psi_n} &= \ket{\Phi_n} + \sum_{A'} \bP_{A',eq}\cdot \ket{\partial_{\bP_A'} \Phi_n}\\
    &= \ket{\Phi_n} + \sum_{A'} \bP_{A',eq} \cdot \sum_{c\neq n}\ket{\Phi_c}\frac{\bra{\Phi_c}\frac{\partial \hH_{BH}}{\partial \bP_{A'}}\ket{\Phi_n}}{E_c-E_n}
\end{align}
Here $\hH_{BH}$ is the semiclassical Born-Huang Hamiltonian (popularized by Shenvi\cite{Shenvi2009-jcp}), such that the equilibrium derivative is $\partial_\bP \hH_{BH} = -i\hbar \hat{\bd}^A M_A^{-1}$ where  $\hat{\bd}^A$ is the derivative coupling. To derive Nafie's equality, note that:
\begin{align}
    \frac{\partial}{\partial\bP_A}\bra{\Psi_n}\bhp\ket{\Psi_n} &= \frac{\partial}{\partial\bP_A}\left( \sum_{A'} \frac{2\hbar}{M_A'} \mathrm{Im}\left[\sum_{c\neq n} \bra{\Phi_n}\bhp\ket{\Phi_c} \frac{\bra{\Phi_c}\bP_{A'}\cdot \hat{\bd}^{A'}\ket{\Phi_n}}{E_c-E_n}\right]\right) \\
    &=\frac{2 m_e}{M_A} \mathrm{Re}\left[ \sum_c \bra{\Phi_n}\bhr\ket{\Phi_c} \bra{\Phi_c}\hat{\bd}^A\ket{\Phi_n}\right]
\end{align}
Here we have used the time independent fact that $-\frac{i\hbar}{m_e}\bra{\Phi_c}\bhp\ket{\Phi_n} = (E_c-E_n)\bra{\Phi_c}\bhr\ket{\Phi_n}$ and taken the $\bP$ derivative. Lastly, we remove the identity above $\left(\sum_c \ket{\Phi_c}\bra{\Phi_c}\right)$ and we recall the standard definition of the derivative coupling for molecules, $\bra{\Phi_c}\hat{\bd}^A\ket{\Phi_n} = \braket{\Phi_c}{\partial_{\bR_
A} \Phi_n}$. The result is:
\begin{align}
    \frac{\partial}{\partial\bP_A}\bra{\Psi_n}\bhp\ket{\Psi_n} &= 2 \frac{m_e}{M_A} \mathrm{Re}\left[ \bra{\Phi_n}\bhr\ket{\partial_\bR\Phi_n}\right] \\
    &=\frac{m_e}{M_A} \frac{\partial}{\partial \bR_A}\bra{\Phi_n}\bhr\ket{\Phi_n}
\end{align}
which proves Eq. \ref{eq:nafie_molecular} above.

\subsection{The Nafie Equality for Periodic Systems }\label{sec:nafie_periodic}

Let us now turn to the periodic case. Immediately,  differences arise from the molecular case. First,
 within the typical one-electron BO picture of solid state systems, the left  hand side of Eq. \ref{eq:nafie_molecular} is fraught: $\langle \br\rangle$ cannot be well defined within one unit cell; at the same time, of course, the right hand side remains difficult because, just as for molecules, there is no existing notion of $\bP_A$ within a conventional  BO picture. Second, for a solid (unlike a molecule), all atomic derivatives $\partial_{\bR_A}$ needs to be redefined in terms of a fourier component with a wave vector, e.g. $\partial_{\bQ_{A\bq}}$, as in the `monochromatic'  perturbation theory for phonon structure discussed above.\cite{Stengel2026} 

Notwithstanding these conceptual problems, \jln{if $\hA$ is an electronic operator of interest that does not commute with $\hbr$,  as discussed in the main text, it is common to define a $\bq-$dependent operator
\begin{align}
    \hA_{\bq} = \frac{1}{2}\left(e^{-i\bq \cdot \hbr} \hA + \hA e^{-i\bq \cdot \hbr} \right)
\end{align}
with the property that
\begin{align}
\hA_{\bq}^{\dagger}  = \hA_{-\bq}
\end{align}
just as for $Q_{A \bq}$, etc. Below, we will work with the operator $\hat{p}_{\bq}$.  Moreover, let us define the (scalar) operator
\begin{align}
    \zeta_{\bq} =  \left( \frac{1 - e^{-i \bq \cdot \hbr}}{i|\bq|} \right) 
\end{align}
 Note that, as $\bq \rightarrow 0$, $\zeta_{\bq} \rightarrow \bm{n}_\bq \cdot \hbr$,  where $\bm{n}_\bq = \bq/|\bq|$ is the unit-vector along $\bq$, so that $\zeta$ is a $\bq$-dependent generalization of the position operator. 

At this point, we submit that the proper Nafie-analogous relationship is:
\begin{equation}
    \mbox{Re} \left\{ \frac{\partial\langle \psi_{n\bk} | \hat \zeta_{\bq}  | \psi_{n\bk} \rangle}{\partial \bQ_{A\bq}} \right\}= \frac{M_A}{m_e}\mbox{Re} \left\{\frac{\partial \bra{\psi_{n\bk}} \hat \bp_{\bq} \cdot  \bm{n}_\bq\ket{\psi_{n\bk}}}{\partial \bPi_{A,-\bq}}\right\}
    \label{eq:analogue_SM}
\end{equation}
 
Eq. \ref{eq:analogue_SM} reduces to the standard molecular Nafie relationship, albeit in one dimension. 
We emphasize that this relationship  exists only in one-dimensional slices when $\bq \ne 0$: $\zeta_{\bq}$  is not really a vector, but rather a scalar along the same $\bq$ direction as the phonon.
}

For use below, we will also define mixed nuclear-electronic curvature-like tensors, where $\alpha$ is a nuclear coordinate that depends on $\bq$ and $\beta$ is an electric operator:
\begin{align}
    \Omega^{\bq}_{n\bk}(\alpha_{\bq},\beta) &\equiv - \frac{1}{2}
    \mbox{Re} \left( \left<
    \frac{\partial}{\partial \alpha_{-\bq}} \psi_{n \bk} \middle| i e^{-i\bq\cdot\br}  \frac{\partial}{\partial \beta} \psi_{n \bk} 
    \right>
    -
    \left<
    \frac{\partial}{\partial \beta} \psi_{n \bk}
     \middle| i e^{-i\bq\cdot \br}  
     \frac{\partial}{\partial \alpha_{\bq}} \psi_{n \bk}
    \right>
    \right) \\
    &= 
     \frac{1}{2}
    \mbox{Im} \left( \left<
    \frac{\partial}{\partial \alpha_{-\bq}} \psi_{n \bk} \middle|  
    e^{-i\bq\cdot\br}  \frac{\partial}{\partial \beta} \psi_{n \bk} 
    \right>
    -
    \left<
    \frac{\partial}{\partial \beta} \psi_{n \bk}
     \middle| 
     e^{-i\bq\cdot \br}  
     \frac{\partial}{\partial \alpha_{\bq}} \psi_{n \bk}
    \right>
    \right)
\end{align}
Lastly, let us also note that, for a periodic system, the proper resolution of the identity is:

\begin{align}
    \hat{I} = \frac{1}{N} \sum_{\bk, p} \ket{\psi_{p\bk}}
    \bra{\psi_{p\bk}} 
\end{align}

\subsubsection{Evaluation of the LHS \label{sec:Nafie_LHS}} 
As is well-known in the periodic solid state world, defining the electronic position operator is difficult for periodic systems. The usual resolution to this paradox is the Berry phase formalism. Here, we start with the identity
\begin{align}
    \left[ \hH_W, \bhr\right] = \frac{-i\hbar}{m_e} \bhp,
\end{align}
from which
it follows that (assuming $\epsilon_{m\bk'} \ne \epsilon_{n\bk}$)
\begin{align}
    \bra{\psi_{m\bk'}} \hat \br\ket{\psi_{n\bk}} = \frac{i \hbar}{m_e} \frac{\left< \psi_{m \bk'} \middle| \bhp \middle| \psi_{n\bk} \right>}{\epsilon_{n\bk} - \epsilon_{m \bk'}}
    =
    \frac{i \hbar}{m_e} \frac{\left< u_{m \bk'} \middle| ( \bhp + \hbar \bk) \middle| u_{n\bk} \right>}{\epsilon_{n\bk} - \epsilon_{m \bk'}}\delta_{\bk,\bk'}
    \label{eq:defr1}
\end{align}
While Eq. \ref{eq:defr1} defines $\bhr$ inside of one $k-$point, this definition can be naturally generalized to include $q-$ variations along the direction of $\bq$ as follows (by replacing $\br$ with $\zeta_{\bq}$ and $\bp$ with $\bp_{\bq}$):
\begin{align}
    \bra{\psi_{m\bk'}} \hat{\zeta}_{\bq}\ket{\psi_{n\bk}} \equiv 
     \frac{i \hbar}{m_e} \frac{\left< \psi_{m \bk'} \middle| \bhp_{\bq} \cdot \bm{n}_\bq \middle| \psi_{n\bk} \right>}{\epsilon_{n\bk} - \epsilon_{m \bk'}}= 
    \frac{i \hbar}{m_e} \frac{\left< u_{m \bk'} \middle| \jln{\bm{n}_\bq \cdot(\bhp + \hbar\bk) - \hbar\bq/2} \middle| u_{n\bk} \right>}{\epsilon_{n\bk} - \epsilon_{m \bk'}}\delta_{\bk-\bk',\bq} \label{eq:def_<zetaq>}
\end{align}
\jln{ Additionally, we note that $\zeta_\bq$ and $\bhp_\bq\cdot \bm{n}_\bq$ have the same canonical relationship to $\hH_W$ as $\bhr$ does to $\bhp$, which enables the evlaution in Eq. \ref{eq:def_<zetaq>}
\begin{align}
    \left[\hH_W, \zeta_\bq\right] &= -i\hbar \frac{\bhp_\bq\cdot \bm{n}_\bq}{m_e}. 
\end{align}}

Note that, for one $k-$point, we can write Eq. \ref{eq:defr1} above as
\begin{align}
    \bra{\psi_{m\bk'}} \hat \br\ket{\psi_{n\bk}} = 
    {i} \frac{\left< u_{m \bk'} \middle| \partial H_k/\partial \bk \middle| u_{n\bk} \right>}{\epsilon_{n\bk} - \epsilon_{m \bk'}}\delta_{\bk,\bk'}
    = i
    \left< u_{m \bk'} \middle| \frac{\partial}{ \partial \bk}  u_{n\bk} \right>,\label{eq:berry_connection}
\end{align}
This expression leads to the notion that $\hbr \rightarrow i \partial/\partial \bk$ within one $k$-point.  Thus, we can similarly think of  Eq. \ref{eq:def_<zetaq>} as leading a heuristic definition $\hat{\zeta}_{\bq} \rightarrow i e^{-i \bq \cdot \br} \partial/\partial \tilde{\bk}$, though this ``definition'' of $\tilde{\bk}$ is no more than shortcut for Eq. \ref{eq:def_<zetaq>} when $\bq \neq 0$.

Before we compute the derivative, \jln{note that the LHS of Eq. \ref{eq:analogue_SM} can be put in an intuitive Berry-curvature  form}:
\begin{align}
\mbox{Re} \left\{ \frac{\partial\langle\psi_{n\bk}| \hat{\zeta}_{\bq} |\psi_{n\bk} \rangle} {\partial \bQ_{A\bq}}\right\}
&= \mbox{Re} \Biggl( \braket{\nabla_{\bQ_{A-\bq}} \psi_{n\bk}}{\zeta_{\bq} \psi_{n\bk}} + \braket{ \psi_{n\bk}}{\zeta_{\bq} \nabla_{\bQ_{A\bq}} \psi_{n\bk}} \Biggr)
\\
&= - \mbox{Im} \Biggl( \braket{\nabla_{\bQ_{A -\bq}} \psi_{n\bk}}{  e^{-i\bq \cdot \br} \frac{\partial}{\partial \tilde{\bk}}   \psi_{n\bk}} - 
\braket{  \frac{\partial}{\partial \tilde{\bk}}   \psi_{n\bk}}{  e^{-i\bq \cdot \br}  \nabla_{\bQ_{A \bq}} \psi_{n\bk}}
\Biggr)
\\
 &= - 2 \Omega_{n\bk}^{\bq}(\bQ_{A\bq},\tilde{\bk})
 \label{eq:berry1}
\end{align}

Finally, we evaluate the response terms by inserting a resolution of the identity
\begin{align}
    \mbox{Re} \left\{\frac{\partial\langle\psi_{n\bk}| \zeta_\bq |\psi_{n\bk} \rangle}{\partial \bQ_{A\bq}}  \right\}
    & = 
    -\frac{1}{N} \sum_{\bk' c} \Re{\braket{ 
    \nabla_{\bQ_{A,-\bq}}
     \psi_{n\bk}}
    {\psi_{c\bk'}}
    \braket{\psi_{ck'}}{
    \zeta_\bq \psi_{n\bk} 
     }}
      \nonumber \\
     &+ 
      \frac{1}{N} \sum_{\bk' c} \Re{\braket{\zeta_{-\bq} \psi_{n\bk}}{\psi_{c\bk'}}\braket{\psi_{ck'}}{ 
    \nabla_{\bQ_{A\bq}}
     \psi_{n\bk}}}
\end{align}
and using Eqs. \ref{eq:dQpsi2} and \ref{eq:def_<zetaq>} to find:  
\jln{
\begin{align}
\mbox{Re} \left\{ \frac{\partial\langle\psi_{n\bk}| \hat \zeta_\bq |\psi_{n\bk} \rangle}{\partial \bQ_{A\bq}} \right\}
    &=  
       - N  \sum_{\bk',c\neq n}  
\Im{\frac{\bra{u_{n\bk}} 
\delta \hat{W}^{A\bq}
    \ket{u_{c \bk'}}\bra{u_{c \bk'}}
\bm{n}_\bq \cdot(\bhp + \hbar\bk) - \hbar\bq/2
    \ket{u_{n\bk}}}{(\epsilon_{c \bk'}-\epsilon_{n \bk})^2}} 
\delta_{\bk',\bk-\bq}
\nonumber
   \\
   & +
 N  \sum_{\bk',c\neq n}  
\Im{\frac{\bra{u_{n\bk}} 
\bm{n}_\bq \cdot(\bhp + \hbar\bk) - \hbar\bq/2
    \ket{u_{c \bk'}}\bra{u_{c \bk'}}
    \delta \hat{W}^{A\bq}
    \ket{u_{n\bk}}}{(\epsilon_{c \bk'}-\epsilon_{n \bk})^2}} 
\delta_{\bk',\bk+\bq}
\label{eq:nafie_RHS} 
\end{align}
}

In the limit that $\bq \to 0$, we find:
\begin{align}
\frac{\partial\langle\psi_{n\bk}| \hat \br |\psi_{n\bk} \rangle}{\partial \bQ_{A,\bq=0}}
    &=  
        2N  \sum_{c\neq n}  
\Im{\frac{\bra{u_{n\bk}} 
\jln{\bhp + \hbar \bk}
    \ket{u_{c \bk}}\bra{u_{c \bk}}
\delta \hat{W}^{A,\bq = 0}
    \ket{u_{n\bk}}}{(\epsilon_{c \bk}-\epsilon_{n \bk})^2}}  
\end{align}

\subsubsection{Evaluation of the RHS\label{sec:Nafie_RHS}} 
To evaluate the right hand side of Eq. \ref{eq:analogue_SM}, we require a Hamiltonian parametrized by both $\bQ$ and $\bPi$ in the spirit of Shenvi's approach.  This Hamiltonian is clearly   given by $\hat{H}'_W$ in Eq. \ref{eq:SM:H_QShenvi},
which satisfies (for $n \ne m$):

\begin{align}
    \left.\left(\frac{\partial H'_W}{\partial \bPi_{A,-\bq}} \right)_{n\bk',m\bk}\right|_{\bPi=0} &= \left.\left(\left(\frac{\bPi_{A\bq}}{M_A}\right)_{n\bk',m\bk}  - \frac{i\hbar}{{M_A}} \bd^{A\bq}_{n\bk',m\bk}\right)\right|_{\bPi=0} =  \frac{-i\hbar}{{M_A}} \bd^{A \bq}_{n\bk'm\bk}\label{eq:dHdp_ad}
\end{align}

Before we evaluate the LHS of Eq. \ref{eq:analogue_SM}, let us relate this matrix element to the relevant curvature tensor.
 Recalling the Hellman-Feynman definition of the derivative coupling in Eq. \ref{eq:dAq}, the LHS can be written down as:
\begin{align}
     \frac{M_A}{me}\mbox{Re} \left\{ \frac{\partial \bra{\psi_{n\bk}} \jln{\bhp_\bq\cdot\bm{n}_\bq} \ket{\psi_{n\bk}}_{\bq}}{\partial \bPi_{A,-\bq}} \right\}
& =   \frac{M_A}{m_e} \Re \left( \left<\partial_{\bPi_{A\bq}} \psi_{n\bk} \middle| \jln{\bhp_\bq\cdot\bm{n}_\bq} \;     \psi_{n\bk} \right> + 
    \left< \jln{\bhp_{-\bq}\cdot\bm{n}_{\bq}} \psi_{n\bk} \middle|       \partial_{\bPi_{A- \bq}}   \psi_{n\bk} \right>
    \right) 
\end{align}
Now, consider the second term on the RHS above, and note that:
\begin{align}
\left< \jln{\bhp_{-\bq}\cdot\bm{n}_{\bq}} \psi_{n\bk} \middle|       \partial_{\bPi_{A-\bq}}   \psi_{n\bk} \right> = 
\left< \jln{\hbar \bq  e^{i\bq\cdot\bhr}} \psi_{n\bk} \middle|       \partial_{\bPi_{A-\bq}}   \psi_{n\bk} \right> + 
\left< \jln{ e^{i\bq\cdot\bhr} (\bhp \cdot \bm{n}_\bq)}  \psi_{n\bk} \middle|       \partial_{\bPi_{A-\bq}}   \psi_{n\bk} \right>
\end{align}
By Eq. \ref{eq:forbelow} above and intermediate normalization for the perturbation, it follows that $\left<    e^{i\bq\cdot\bhr}\psi_{n\bk} \middle|       \partial_{\bPi_{A-\bq}}   \psi_{n\bk} \right> = 0 $ \jln{so that the first term above vanishes and 
therefore,} 
\begin{align}
\frac{M_A}{me}\mbox{Re} \left\{\frac{\partial \bra{\psi_{n\bk}} \jln{\bhp_\bq\cdot\bm{n}_\bq} \ket{\psi_{n\bk}}_{\bq}}{\partial \bPi_{A,-\bq}}   \right\}& =
\frac{M_A}{m_e} \Re \left( \left<\partial_{\bPi_{A \bq}} \psi_{n\bk} \middle| \jln{e^{-i\bq\cdot\bhr}\bhp \cdot\bm{n}_\bq}\;     \psi_{n\bk} \right> + 
    \left< \jln{e^{i\bq\cdot\bhr} \bhp\cdot\bm{n}_\bq} \psi_{n\bk} \middle|       \partial_{\bPi_{A- \bq}}   \psi_{n\bk} \right>
    \right) \\
  & =  - \frac{M_A \hbar }{m_e} \Re \left( \left<\partial_{\bPi_{A \bq}} \psi_{n\bk} \middle| i \jln{e^{-i\bq\cdot\br}} \partial_\br \;     \psi_{n\bk} \right> - 
    \left< \partial_\br   \psi_{n\bk} \middle|   i \jln{e^{-i\bq\cdot\br}}   \partial_{\bPi_{A-\bq}}   \psi_{n\bk} \right>
    \right) 
    \\
      & =  \frac{M_A \hbar }{m_e} \Im \left( \left<\partial_{\bPi_{A \bq}} \psi_{n\bk} \middle|  \jln{e^{-i\bq\cdot\br}} \partial_\br \;     \psi_{n\bk} \right> - 
    \left< \partial_\br   \psi_{n\bk} \middle|    \jln{e^{-i\bq\cdot\br}}   \partial_{\bPi_{A-\bq}}   \psi_{n\bk} \right>
    \right) 
    \\
    &= \frac{2M_A\hbar}{m_e} \Omega^{\bq}_{n\bk,n\bk}(\bPi_{A-\bq}, \br)
    \label{eq:berry2}
\end{align}

Finally, let us  compute the matrix element derivative and establish Nafie's equality in a solid:
\jln{
\begin{align}
\nonumber
    & \frac{M_A}{m_e}\mbox{Re} \left\{ \frac{\partial \bra{\psi_{n\bk}} \bhp_\bq\cdot\bm{n}_\bq \ket{\psi_{n\bk}}_{\bq}}{\partial \bPi_{A,-\bq}}\right\} \\
    & \; \; \; \; \; \; \; \; \; \; \; =  \frac{M_A}{m_e} \Re \left( \left<\partial_{\bPi_{A,\bq}} \psi_{n\bk} \middle| \bhp_\bq\cdot\bm{n}_\bq \;     \psi_{n\bk} \right> + 
    \left< \bhp_{-\bq} \cdot\bm{n}_\bq\psi_{n\bk} \middle|       \partial_{\bPi_{A,-\bq}}   \psi_{n\bk} \right>
    \right) \label{eq:dPdp}
    \\
  & \; \; \; \; \; \; \; \; \; \; \; =\frac{1}{N} \sum_{c\bk'}  \frac{M_A}{m_e}\Re \left(\bra{\partial_{\bPi_{A,\bq}} \psi_{n\bk}}  \ket{\psi_{c\bk'}}\bra{\psi_{c\bk'}}  \ket{ \bhp_\bq\cdot\bm{n}_\bq \; \psi_{n\bk}} 
  + 
  \bra{ \bhp_{-\bq}\cdot\bm{n}_\bq
  \psi_{n\bk}}  \ket{\psi_{c\bk'}}\bra{\psi_{c\bk'}}  \ket{  
  \partial_{\bPi_{A-\bq}}
  \; \psi_{n\bk}}
  \right) \label{eq:dPdp2}
    \\
    & \; \; \; \; \; \; \; \; \; \; \; =  \frac{1}{N}  \frac{M_A}{m_e} \sum_{\bk',c\neq n} \Re{\frac{\bra{\psi_{n\bk}} \left(\frac{\partial H'_W}{\partial \bPi_{A,\bq}} \right)^{\dagger} \ket{\psi_{c\bk'}}}{\epsilon_{c\bk'}-\epsilon_{n\bk}}  \bra{\psi_{c\bk'}}
    \ket{\bhp_\bq\cdot\bm{n}_\bq   \psi_{n\bk}} + 
    \bra{\bhp_{-\bq}\cdot\bm{n}_\bq\psi_{n\bk}}    \ket{\psi_{c\bk'}}
    \frac{\bra{\psi_{c\bk'}}\frac{\partial H'_W}{\partial \bPi_{A,-\bq}}\ket{\psi_{n\bk}}}{\epsilon_{n\bk}-\epsilon_{c\bk'}}  } 
    \label{eq:nafieRHS_dHdPi}\\
    & \; \; \; \; \; \; \; \; \; \; \;= \frac{1}{N}  \frac{M_A}{m_e} \sum_{\bk',c\neq n}  \Re{\frac{\bra{\psi_{n\bk}}\frac{i\hbar}{M_A}\hat{d}^{A\bq}\ket{\psi_{c \bk'}}}{\epsilon_{c\bk'}-\epsilon_{n\bk}}\bra{\psi_{c\bk'}} \ket{\bhp_\bq\cdot\bm{n}_\bq \psi_{n\bk}} 
    + 
    \bra{\bhp_{-\bq}\cdot\bm{n}_\bq  \psi_{n\bk}}   \ket{\psi_{c\bk'}}
    \frac{\bra{\psi_{c\bk'}}\frac{-i\hbar}{M_A}\hat{d}^{A\bq}\ket{\psi_{n \bk}}}{\epsilon_{n\bk}-\epsilon_{c\bk'}}
 }\label{eq:nafieRHS_dAq}
    \\
&\; \; \; \; \; \; \; \; \; \; \; = - N  \Im{ \sum_{\bk',c\neq n}\frac{\bra{u_{n\bk}}\delta \hat{W}^{A\bq} \ket{u_{c \bk'}}}{ (\epsilon_{c\bk'}-\epsilon_{n\bk})^2} \left< u_{c\bk'} \middle| \bm{n}_\bq \cdot(\bhp + \hbar\bk) - \hbar\bq/2   \middle| u_{n\bk} \right> 
}   \delta_{\bk',\bk-\bq}  \nonumber
\\
&\; \; \; \; \; \; \; \; \; \; \; \; \; \;
+ N  \Im{ \sum_{\bk',c\neq n} \left< u_{n\bk} \middle| \bm{n}_\bq \cdot(\bhp + \hbar\bk) - \hbar\bq/2  \middle| u_{c\bk'}\right> \frac{\bra{u_{c\bk'}}\delta \hat{W}^{A\bq} \ket{u_{n \bk}}}{ (\epsilon_{c\bk'}-\epsilon_{n\bk})^2} }   \delta_{\bk',\bk+\bq} 
\label{eq:nlrhs}
\end{align}
}

Eq. \ref{eq:nlrhs}  matches Eq. \ref{eq:nafie_RHS} and  gives a unique equality between  curvatures (see Eqs. \ref{eq:berry1} and \ref{eq:berry2}):
\begin{equation}
    \Omega^{\bq}_{n\bk}(\bQ_{A\bq},\tilde{\bk}) = - \frac{M_A \hbar }{m_e} \Omega^{\bq}_{n\bk}(\bPi_{A-\bq}, \br) 
\end{equation}

\subsubsection{Translational Invariance  within Nafie's equality}
At $\bq=0$, there is an additional symmetry worth noting:
\begin{align}
    \sum_A \frac{\partial \langle \bhr\rangle}{\partial \bQ_{A,\bq=0}} = 0.
    \label{eq:zerozero}
\end{align}
Eq. \ref{eq:zerozero} can be  proven by evoking the sum-rule symmetry of the derivative couplings at $\bq=0$ (Eq. \ref{eq:sumdA0}) and  Nafie's relationship that appears in Eq. \ref{eq:nafieRHS_dAq}. Altogether, we find: 
\begin{align}
    \sum_A \frac{\partial \langle \bhr\rangle}{\partial \bQ_{A,\bq=0}} &= \frac{2N}{m_e}\sum_{c\neq n} \mathrm{Im}\left\{ \frac{\bra{u_{n\bk}}\bhp\ket{u_{c\bk}}\bra{u_{c\bk}}\bhp\ket{u_{n\bk}}}{(\epsilon_{c\bk}-\epsilon_{n\bk})^2}\right\} = 0
    \label{eq:mass_but_im}
\end{align}
Clearly, the RHS of Eq. \ref{eq:mass_but_im} is zero.  Interestingly,  the expression inside the curly brackets above is the inverse of the electronic contribution to effective mass of the electron (which is strictly real and not imaginary).

\section{An approximate Phase Space theory for periodic systems}

The ansatz of phase space electronic structure theory is that, rather than diagonalizing the BO electronic Hamiltonian that depends only on $\bQ$, we diagonalize instead:
\begin{align}
    \hat{H}_{PS}(\bQ,\bPi) = \hH_{el}(\bQ) + \sum_{A \bq} \frac{\bPi_{A -\bq} \cdot \bPi_{A \bq}}{2M_A}   - i \hbar  \sum_{A \bq} \frac{\bPi^{ A \bq} \cdot \hat{\bm{\Gamma}}^{A-\bq}}{M_A} + \hat{\zeta}
    \label{eq:HPS:SM}
\end{align}

In electronic index form, this Hamiltonian takes the form: 

\begin{align}
    (\hat{H}^{PS})_{m\bk',n\bk} =  H^{el}_{m\bk',n\bk} + \delta_{mn}\delta_{\bk,\bk'} \left(   \sum_{A \bq} \frac{\bPi_{A,-\bq} \cdot \bPi_{A \bq}}{2M_A} \right)  - i \hbar \frac{\bPi^{A\bq} \cdot \bGm^{A, -\bq}_{m\bk',n\bk}}{M_A} + \zeta_{m\bk',n\bk} \label{eq:H_PS_qdep}
\end{align}
In practice, diagonlizing this $\hbGm$ and $\hat{H}^{el}$ simultaneously is only possible at $\bQ=0$ (where there is periodicity over the unit cell and Bloch's theorem applies). 

Now, for a one-dimensional molecular (non-periodic) problem, it is well known\cite{Bian2026-cpr} that  a reasonable form of $\hat{\bf \Gamma}^{Al}$ for atom $A$ in unit cell $A$ is:

\begin{align}
    \hat{\bm \Gamma}^{Al} 
    = \frac{1}{2i\hbar}   \left( \theta_{A\ell}(\br)    \bhp + \bhp  \theta_{A\ell}(\br) \right) \label{eq:Gamma_Al}
\end{align}
where $\theta_{A\ell}(\br)$ is one element of a partition of unity that is nonzero when $\br$ is close to $\bR_{A\ell}$.

Inspired by Eq. \ref{eq:Gamma_Al}, we have argued in the main letter that in order to form a proper, anti-hermitian, periodic form of $\hbGm$,
we should construct the periodic function 
\begin{align}
        \tilde{\theta}^{A\bq}(\br)  & \equiv  \sum_{\ell}  \theta_{A\ell}(\br)   e^{-i(\br-\bR_{\ell})\cdot\bq}  \label{eq:theta_Aq}
\end{align}

$\tilde{\theta}^{A\bq}(\br)$  is not hermitian but is periodic on the unit cell, which can be proven as follows:
\begin{align}
    \tilde{\theta}^{A\bq}(\br+\ba_1)  & = \sum_{\ell'}  \left( \theta_{A\ell'}(\br+\ba_1)   e^{-i(\br+\ba_1-\bR_{\ell'})\cdot\bq}   \right)  \\
     & = \sum_{\ell}  \left( \theta_{A\ell}(\br)   e^{-i(\br-\bR_{\ell})\cdot\bq}  \right)
     \\
     &= \tilde{\theta}^{A\bq}(\br) \label{eq:Thetaq_periodic}
\end{align}
Here, we have substituted $\ell' = (\ell_1+1, \ell_2, \ell_3)$ and used the fact that
\begin{align}
    \theta_{A \ell'}(\br + \ba_1) = \theta_{A\ell}(\br).
\end{align}
(A similarly applies for shifts on axes $\ba_2$ and $\ba_3$.)

In the main paper, we have further submitted that the proper form for $\hbGm$ over the whole solid is then
\begin{align}
    \hat{\Gamma}^{A\bq}  & =  \jln{\frac{1}{2} \left(e^{i\bq \cdot \br} \tilde{\Gamma}^{A\bq} \jln{+\tilde{\Gamma}^{A\bq}e^{i\bq\cdot\br}}\right)}
\end{align}
where the operator  over the unit cell is
\begin{align}
    \tilde{\bf \Gamma}^{A{\bq}}  & \equiv  \frac{1}{2i\hbar}  \left( \tilde{\theta}_{A\bq}(\br)  \bhp + \bhp \tilde{\theta}_{A\bq}(\br) \right), 
\end{align}
Like $\tilde{\theta}_{Aq}$,
$\tilde{\bm \Gamma}^{A\bq}$ is also periodic.  The relevant operators satisfy:
\begin{align}
    \left(\tilde{\theta}^{A \bq}\right)^{\dagger} &= \tilde{\theta}^{A -\bq} \\
     \left(\tilde{\bm \Gamma}^{A \bq}\right)^{\dagger} &= -\tilde{ \bm
     \Gamma}^{A -\bq} \\
    \left(\hat{\bm \Gamma}^{A \bq}\right)^{\dagger} &= -\hat{ \bm
     \Gamma}^{A -\bq} 
\end{align}

Lastly, note that we can also write
\begin{align}
\joe{
    \hat{\Gamma}^{A\bq}(\br)   =  \frac{1}{2i\hbar} \left(\sum_{\bl} \theta_{A,l}(\br)  e^{i\bR_l\cdot\bq} \bhp + \bhp e^{i\bR_l\cdot\bq} \theta_{A,l}(\br)  \right)} 
\label{eq:hatGammaq_all}
\end{align}
which is reported in the main text.     

Finally, let us evaluate $\hbGm$ operators in a basis:
\begin{align}
     \bra{\psi_{m\bk'}} \hat{\bm \Gamma}^{A\bq}(\br) \ket{\psi_{n\bk}} &=
     \bra{\psi_{m\bk'}} \jln{\frac{1}{2} \left(e^{i\bq \cdot \br} \tilde{\Gamma}^{A\bq}(\br) \jln{+\tilde{\Gamma}^{A\bq}(\br)e^{i\bq\cdot\br}}\right)} \ket{\psi_{n\bk}}
     \\ &=
    \frac{1}{2}\left( \bra{\psi_{m\bk'}} e^{i\bq\cdot\br}\tilde{\bm \Gamma}^{A\bq}(\br) \ket{\psi_{n\bk}} +\jln{\bra{\psi_{m\bk'}} \tilde{\bm \Gamma}^{A\bq}(\br)e^{i\bq\cdot\br} \ket{\psi_{n\bk}} }\right)
     \\
     &= \jln{\frac{1}{2}}\int d\br e^{-i\bk'\cdot\br} u^*_{m\bk'}(\br) e^{i\bq\cdot\br} \frac{1}{2} \left( \tilde{\theta}_{A\bq}\bhp +\hbp \tilde{\theta}_{A\bq} \right)e^{i\bk\cdot\br}u_{n\bk}(\br)
     \nonumber\\
     &\qquad \jln{+ \frac{1}{2}\int d\br e^{-i\bk'\cdot\br} u^*_{m\bk'}(\br)  \frac{1}{2} \left( \tilde{\theta}_{A\bq}\bhp +\hbp \tilde{\theta}_{A\bq} \right)e^{i\bq\cdot\br}e^{i\bk\cdot\br}u_{n\bk}(\br)} \\
&= N \delta_{\bk',\bk+\bq} \bra{u_{m\bk'}}\frac{1}{2i\hbar}\left( \tilde{\theta}_{A\bq}\bhp + \bhp\tilde{\theta}_{A\bq}  + \hbar \jln{(\bk+\bk')} \tilde{\theta}_{A\bq}\right)\ket{u_{n\bk}}  \\
&\equiv
N \bra{u_{m,\bk+\bq}}\dt{\bm \Gamma}^{A\bq}\ket{u_{n\bk}} \delta_{\bk',\bk+\bq}
\end{align}
where
\begin{align}
\bra{u_{m,\bk+\bq}}\dt{\bm \Gamma}^{A\bq}\ket{u_{n\bk}} &\equiv  \bra{u_{m,\bk+\bq}}\frac{1}{2i\hbar}\left( \tilde{\theta}^{A\bq}(\bhr)\bhp + \bhp\tilde{\theta}^{A\bq}(\bhr) + \jln{\hbar(2\bk+\bq)}\tilde{\theta}^{A\bq}(\bhr)\right)\ket{u_{n\bk}} \label{eq:dt_theta}
\end{align}

Note that $\bra{u_{m,\bk+\bq}}\dt{\bm \Gamma}^{A\bq}\ket{u_{n\bk}} \ne \bra{u_{m,\bk+\bq}}\tilde{\bm \Gamma}^{A\bq}\ket{u_{n\bk}}$ because of the presence of the momentum operator $\bhp$ in the definition of $\hat{\bf \Gamma}^A$.

\subsection{A Plane Wave Basis}

In general, if we wish to expand $\psi_{nk}$ in a plane-wave basis, the proper basis is of the form $\ket{k + G}$.
Now, consider the case where we wish to work on a a single unit cell; let us write  $\br \to \bx + \bR_\ell$. 
In a plane wave basis,  we can always write $\braket{\bx}{\bk+\bG} = e^{i\bx\cdot(\bk+\bG)}$.

With this fact in mind, if we wish to evaluate matrix elements of  $\hat{\bm \Gamma}^{A\bq}$, a simple calculation shows:
\begin{align}
    \bra{\bk+\bG+\bq} \hat{\bm \Gamma}^{A\bq}\ket{\bk+\bG'} &= 
    \bra{\bk+\bG} \tilde{\bm \Gamma}^{A\bq}\ket{\bk+\bG'}
    \\
    &= \frac{-iN}{2}\tilde{\theta}_{A\bq}^f(\bG'-\bG)(\bG'+\bG + \jln{2\bk+\bq})  
\end{align}

\begin{align}
\tilde{\theta}_{A\bq}^f(\bG'-\bG) & = \int d\bx \, \tilde{\theta}_{A\bq}(\bx)e^{i (\bG' - \bG) \cdot \bx} \\
& = \sum_{\ell} \int d\bx   {\theta}_{A\ell}(\bx)
e^{-i \bq \cdot (\bx - \bR_{\ell})}
e^{i (\bG' - \bG) \cdot \bx} \\
& = \sum_{\ell} \int d\bx   {\theta}_{A\ell = 0}(\bx + \bR_{\ell})
e^{i (\bG' - \bG -\bq) \cdot (\bx - \bR_{\ell})}
\label{eq:fromone}
\end{align}
Above, we have used the identity that $\bG \cdot  \bR_{\bl} = 2 \pi n$ for $n \in \mathbb{Z}$.  If one would compare the notation to say Ref. \cite{Baroni2001_DFPT}, one might equivalently write the Fourier transform as $\tilde{\theta}_{A\bq}^f(\bG'-\bG) \equiv \tilde{\theta}_{A}^f(\bG'-\bq-\bG)$. 

According to Eq. \ref{eq:fromone}, one need only compute the Fourier transform for $\theta_{A, \ell = 0}$ in order to retrieve $\tilde{\theta}_{A\bq}^f(\bG'-\bG) $ and run a phase space electronic structure calculation. 
For calculations in this paper, we have evaluated matrix elements using the function
\begin{align}
    \theta_{A\ell}(\br) \equiv& \frac{e^{-(\br-\bR_\ell -\tau_A)^2/\sigma^2}}{
\sum_A\sum_\ell e^{-(\br-\bR_\ell-\tau_A)^2/\sigma^2
    } } \label{eq:theta_Al} 
\end{align}
for $\bR_{A\ell}  = \bR_{\ell} + \tau_A$ where $\tau_A$ is the position within the primitive cell of atom $A$.

Before concluding this section, we remind the reader that, when solving a phase space electronic structure problem, one can only work with vibrations at $\bq=0$ if one wishes to maintain periodicity.

\subsection{Momentum Conservation within PS Theory}

Let us now show that dynamics with the Hamiltonian in Eq. \ref{eq:HPS:SM} obey momentum conservation when the nuclei are treated classically. 
The total momentum is  the kinetic energy of the nuclei plus the momentum of the electrons:
\begin{align}
\bP_{tot} &=
    \sum_{A\ell} \left(\bP_{A\ell}  - i\hbar \left<\hat{\bm \Gamma}^{Al}\right> \right)  + \left<\bhp\right> \\
    &=     \sum_{A} \bPi_{A,\bq=0}  - 
    \sum_A i\hbar \left<\hat{\bm \Gamma}^{A,\bq=0}\right>   + \left<\bhp\right>
\end{align}

Now, note the following two identities:
\begin{align}
    -i\hbar \sum_A \hat{\Gamma}^{A,\bq=0} +\bhp   &= 0 \label{eq:Gammaq0_cond1}\\
    \left[ -i\hbar \sum_{A\ell} \nabla_{\bR_{A\ell}} + \bhp, \hat{\Gamma}^{A\bq} \right] = \left[ -i\hbar \sum_A \nabla_{\bQ_{A0}} + \bhp, \hat{\Gamma}^{A\bq} \right] &= 0 \label{eq:Gammaq0_cond2} 
\end{align}
These equations are directly analogous to Eq. \ref{eq:sumdA0} and Eq. \ref{eq:nablaBdAq} for the derivative couplings.

Eq. \ref{eq:Gammaq0_cond1} follows from Eq. \ref{eq:hatGammaq_all} and the fact that at $\bq=0$, $\tilde{\theta}_{A,0}(x)$ is a partition of unity,
\begin{align}
    \sum_A \tilde{\theta}_{A,0}(\br) = \sum_{A\ell} \theta_{A\ell}(\br) =\sum_{A\ell} \frac{e^{-(\br-\bR_\ell -\tau_A)^2/\sigma^2}}{\sum_{A\ell} e^{-(\br-\bR_\ell-\tau_A)^2/\sigma^2
    }} = 1
\end{align}

Eq. \ref{eq:Gammaq0_cond2} can be proven by noting that $\theta_{A\ell}(\br)$ is a function of the position of the electron relative to the nuclei, so that
\begin{align}
    \left[ \sum_{A\ell} \nabla_{\bR_{A\ell}} + \frac{\partial}{\partial \br}, \hat{\theta}^{Al}(\br) \right] &= 0  
\end{align}

Finally, it is always true (also by translational invariance) that:
\begin{align}
    \left[ -i\hbar \sum_A \nabla_{\bQ_{A0}} + \bhp, \hat{H}_{el}\right] &= 0 
    \label{eq:stupid}
\end{align}

so that the PS Hamiltonian is also translationally invariant:
\begin{align}
    \left[ -i\hbar \sum_A \nabla_{\bQ_{A0}} + \bhp, \hat{H}_{PS}\right] &= 0 
    \label{eq:stupid2}
\end{align}

Classical dynamics along an eigenvalue of the phase-space electronic Hamiltonian  in Eq. \ref{eq:HPS:SM} then satisfy momentum conservation because (from Eq. \ref{eq:Gammaq0_cond1}) $\bP_{tot}=
    \sum_{A\ell} \bP_{A\ell} = \sum_A \bPi_{A0}$.
Furthermore, because of Eq.  \ref{eq:stupid2}, it follows that
\begin{align}
    \sum_A \dot{\bPi}_{A0} =  - \sum_A \left< \frac{\partial \hH_{PS}}{\partial Q_{A0}} \right>   =  \frac{1}{i\hbar} \left< \left[ \hH_{PS},\bhp  \right] \right>  = 0
\end{align}


.

\subsection{Verifying Nafie's equality for Phase Space Calculations} 

In Fig. 1  of the paper, we have evaluated Nafie's equality numerically using our phase space electronic Hamiltonian. To make this plot,  we have calculated $\langle \bhp\rangle$ with a phase space electronic structure calculation just as in Sec. \ref{sec:Nafie_RHS} above. 
Note that (for $n \ne m$):
\begin{align}
    \left(\frac{\partial H_{PS}}{\partial \bPi_{A,-\bq}}\right)_{nk',mk} = \left.\frac{1}{M_A}\left( \left( \bPi_{Aq}\right)_{nk',mk} - i\hbar \bra{\psi_{n \bk'}}\hat{\Gamma}^A_{\bq\bk}\ket{\psi_{m\bk}}\right)\right|_{eq} = - \frac{i\hbar N}{M_A}\bra{u_{n \bk'}}\dt{\Gamma}^A_{\bq}\ket{u_{m\bk}}\delta_{\bk',\bk+\bq}\label{eq:dHdPi_PS}
\end{align}

Therefore, we can evaluate the LHS of Nafie's equality with only a first order perturbation theory, starting with Eq. \ref{eq:nafieRHS_dHdPi} but substituting in the derivative in Eq. \ref{eq:dHdPi_PS}:
\jln{
\begin{align}
    \frac{M_A}{m_e}\mbox{Re} \left\{\frac{\partial\langle \hat \bp_{\bq} \cdot  \bm{n}_\bq \rangle}{\partial \bPi_{A,-\bq}} \right\}&=\frac{M_A}{m_e}  \frac{1}{N} \sum_{\bk',c\neq n} \Re{\frac{\bra{\psi_{n\bk}}
    \left(\frac{\partial H'_W}{\partial \bPi_{A\bq}}\right)^{\dagger}
    \ket{\psi_{c\bk'}}}{\epsilon_{c\bk'}-\epsilon_{n\bk}}  \bra{\psi_{c\bk'}}
    \ket{ \hat \bp_{\bq} \cdot  \bm{n}_\bq  \psi_{n\bk}} + 
    \bra{\hat \bp_{-\bq} \cdot  \bm{n}_\bq \psi_{n\bk}}    \ket{\psi_{c\bk'}}
    \frac{\bra{\psi_{c\bk'}}\frac{\partial H'_W}{\partial \bPi_{A,-\bq}}\ket{\psi_{n\bk}}}{\epsilon_{n\bk}-\epsilon_{c\bk'}}  }
    \\
      & = - N \frac{\hbar}{m_e} \sum_{\bk',c\neq n} \Im{\frac{\bra{u_{n\bk}}\dt{\Gamma}^{A,\bq} \ket{u_{c \bk'}}\bra{u_{c\bk'}}\bm{n}_\bq \cdot(\bhp + \hbar\bk) - \hbar\bq/2  \ket{u_{n\bk}}}{\epsilon_{c\bk'}-\epsilon_{n\bk}}} \delta_{\bk',\bk-\bq} \label{eq:Nafie_LHS_PS_SM}\\
    &  +N \frac{\hbar}{m_e} \sum_{\bk',c\neq n} \Im{\frac{\bra{u_{n\bk}}
    \bm{n}_\bq \cdot(\bhp + \hbar\bk) - \hbar\bq/2 
    \ket{u_{c \bk'}}\bra{u_{c\bk'}}
    \dt{\Gamma}^{A,\bq}
    \ket{u_{n\bk}}}{\epsilon_{n\bk}-\epsilon_{c\bk'}}} \delta_{\bk',\bk+\bq}  \nonumber
\end{align}
}
Here, the matrix elements of $\dt{\Gamma}^{Aq}$ are given  in Eq. \ref{eq:dt_theta} above. Eq. \ref{eq:Nafie_LHS_PS_SM} is evaluated and compared against \ref{eq:nlrhs} in Fig. 1 of the main text. The agreement is quite strong, highlighting how and why phase space electronic structure theory is appropriate for and offers new information about solid state materials.

\section{ 1-dimensional Example}
For Figs. 1 and 2 of the main body of the paper, 
we have modeled a one-dimensional system with two nuclei and one electron in a box. The electronic nuclear system is defined with a Lorentzian potential,
\begin{equation}
    V(x) = \sum_{C=\{A,B\}}\sum_{R_\ell} \frac{Z_C}{(x-R_\ell-r_C)^2 + g^2}\Big( 1- (1-2\delta_{BC}) \rm{erf}(\rho (x-R_\ell-\tau_C))\Big)
\end{equation}
where $\rho$ is the skew parameter inverse length. For $\rho=0$, the BO ground state of this system is a Kronig-Penny like metal; for $\rho\neq0$, one recovers a semiconducting (no-degeneracy) state. The data presented in the figures assumes 200 $k$-points and 400 grid points in the unit cell for full convergence. The Lorentzian form of the potential is specifically chosen such that it decays to numerical zero by the sampled 200 unitcells (as compared to a $r^{-1}$ potential). The use of the broadening $g$ parameter is motivated by previous one-dimensional electronic structure calculations.\cite{Helbig2011,Wagner2012,Baker2015} Lastly, we note that nuclear masses are not needed for Figure 2 of the main text because we use velocities $\bP/M$.

\begin{table}[]
    \centering
    \begin{tabular}{|c|c|c|c|c|c|c|}
    \hline
        Fig. &$Z_A$ &  $Z_B$& $g$ & $a$  & $\rho$ & $\sigma$\\
        \hline
         1 & 2 & 2 & 1 & 5.37 & 1 & 2\\
         2 & 2  & 2 & 1 & 5.37 & 0 & 0.5\\
         \hline
    \end{tabular}
    \caption{Parameters for potential, unit cell length, and $\sigma$ for $\hat{\theta}$ for the figures in the main text.}
    \label{tab:placeholder}
\end{table}

\bibliography{main.bib}